\documentclass[12pt, draftclsnofoot, onecolumn]{IEEEtran}
\ifCLASSINFOpdf
\else
\fi

\usepackage{graphicx}
\usepackage[cmex10]{amsmath}
\usepackage{dsfont}
\usepackage{bm}
\usepackage{amssymb}
\usepackage{caption}
\usepackage{array}
\usepackage{algorithm}
\usepackage{algorithmic}
\usepackage{subfigure}
\usepackage{graphics}
\usepackage{epsfig}
\usepackage{float}
\usepackage{extarrows}

\begin{document}
\title{Super-Resolution mmWave Channel Estimation using Atomic Norm Minimization}

\author{Hongyun~Chu,~\IEEEmembership{Student Member,~IEEE,}
        Le~Zheng,~\IEEEmembership{Member,~IEEE,}
        and~Xiaodong~Wang,~\IEEEmembership{Fellow,~IEEE}% <-this % stops a space
\thanks{H. Chu is with the Wireless Energy and Information Transmission Lab, Shenzhen Institutes of Advanced Technology, Chinese Academy of Sciences, Shenzhen, 518055, China.}% <-this % stops a space
\thanks{L. Zheng and X. Wang are with the Electrical Engineering Department, Columbia University, New York 10027, USA, e-mail: le.zheng.cn@gmail.com;
wangx@ee.columbia.edu.}% <-this % stops a space
%\thanks{TCOM version based on Michael Shell's bare{\textunderscore}jrnl.tex version 1.3.}
}

%% The paper headers
%\markboth{IEEE Transactions on Wireless Communications}%
%{Submitted paper}

\maketitle
\begin{abstract}
%\boldmath
We propose super-resolution MIMO channel estimators for millimeter-wave (mmWave)
systems that employ hybrid analog and digital beamforming and generalized spatial modulation, respectively. Exploiting the inherent sparsity of mmWave channels, the channel estimation problem is formulated as an atomic norm minimization that enhances sparsity in the continuous angles of departure and arrival. Both pilot-assisted and data-aided channel estimators are developed, with the former one formulated as a convex problem and the latter as a non-convex problem. To solve these formulated channel estimation problems, we develop a computationally efficient conjugate gradient descent method based on non-convex factorization which restricts the search space to low-rank matrices. Simulation results are presented to illustrate the superior channel estimation performance of the proposed algorithms for both types of mmWave systems compared to the existing compressed-sensing-based estimators with finely quantized angle grids.
\end{abstract}

%\vfill
% \fi
\IEEEpeerreviewmaketitle

\section{Introduction}

\IEEEPARstart{T}{he} millimeter-wave (mmWave) spectrum band has been made available for future 5G wireless communication systems
\cite{5G-1}. To compensate for the severe signal propagation loss at mmWave band and to achieve data rates on the order of gigabits per second, the mmWave systems are expected to employ large antenna arrays at transceivers to provide sufficient beamforming gains. However, due to the high cost of mmWave radio frequency (RF) units, it is infeasible to associate an RF chain with each antenna element in such large mmWave MIMO systems in practice. Two large MIMO architectures have been proposed for mmWave systems that employ a much smaller number of RF chains than the number of antennas at the transceivers. One is the hybrid beamforming (HB) system \cite{Hybrid-MIMO,OMP}, shown in Fig. 1(a), that employs a cascade of analog and digital beamformers at both the transmitter and receiver. The other is the generalized spatial modulation
(GSM) system \cite{GSM-2}, shown in Fig. 1(b), that only activates antennas that are linked to RF chains. Note that in the HB system information bits are carried on the transmitted modulation symbols, whereas in the GSM system information bits are conveyed by both the index set of the activated transmit antennas and the transmitted modulation symbols.
It is well known that in MIMO systems the channel state information (CSI) is indispensable for reliable  signal transmission and reception, and especially useful for designing efficient beamformers in mmWave band \cite{Beamforming}. However, channel estimation is challenging for mmWave systems with a large number of antennas, and
conventional channel estimation methods based on the rich scattering assumption developed for microwave systems
are rather inefficient due to high training overhead and high computational cost.

In \cite{Heath}, it is pointed out that the parametric channel model for mmWave systems leads to a sparse representation of the MIMO channel, which can be exploited for channel estimation purpose - i.e.,  instead of estimating the full channel matrix, one would estimate only the angles of departure/arrival (AoD/AoA) of dominant paths and the corresponding path gains. Leveraging on this, various channel estimators have been proposed in \cite{Beamtraining,Heath,MUSIC,Ziyu,Pilot-reduce,Pilot-reduce2,Angle1,Angle2,OMP} that capitalize on the spatial sparsity of mmWave channels. In particular, in \cite{Beamtraining,Heath}, closed-loop beam training based methods such as multistage beam search are proposed for channel estimation. While such closed-loop methods have been adopted in practical systems, their performance tends to be limited by the training beam patterns.

As an alternative to the closed-loop based beam training techniques, the open-loop techniques perform explicit channel estimation using MUSIC and compressive sensing (CS) methods, by transmitting pilot symbols. In \cite{MUSIC}, a subspace-based mmWave channel estimation method that makes
use of the MUSIC algorithm is proposed. A two-dimensional (2D) MUSIC algorithm for beamformed mmWave MIMO
channel estimation is proposed in \cite{Ziyu}. The MUSIC algorithm is able to identify multiple paths with high resolution but it is sensitive to antenna
position, gain, and phase error. On the other hand, a number of CS-based channel estimators \cite{Pilot-reduce,Pilot-reduce2,OMP,Angle1,Angle2} have been proposed based on the virtual angular domain representation of MIMO channels \cite{Angle1,Angle2},
which describes the channel with respect to some fixed basis functions of angles whose resolution is determined by the spatial resolution of arrays.
%In \cite{OMP}, the authors formulate a sparse signal recovery problem based on the parametric channel model with quantized AoDs/AoAs, called as angle grids, and develop a grid-based orthogonal matching pursuit (OMP) algorithm for estimating channels. The pilot symbols of the proposed schemes are generated by the cascade of the RF beamformer and the baseband MIMO processor.
%Moreover, the massive MIMO channel estimation
%problem is also treated using the CS-based methods in \cite{Est1,Pilot-reduce}.
%These CS-based methods regard the sparse multipath channel
%estimation as a sparse recovery problem. The virtual angular
%representation \cite{Est1,Est3} or angle grids \cite{OMP} are used to describe
%the path directions. For each angle grid that represents a pair
%of transmitting/receiving directions, there is a gain coefficient. Because of
%the spatial sparsity of the mmWave channel, most of the gain
%coefficients should be zeros, which means there is no path
%coming from/to those directions. The sparse recovery method
%aims to recover the nonzero gain coefficients, which naturally
%provides the estimates of the path directions, and at the same
%time reduces the training overhead as much as possible.
%In \cite{Gao}, an adaptive
%structured subspace pursuit algorithm at the user is proposed to jointly estimate channels associated with multiple orthogonal frequency division multiplexing symbols from the limited number of pilots, whereby the spatio-temporal common sparsity of MIMO channels is exploited to improve the channel estimation accuracy.

Different from the traditional grid-based CS techniques, a gridless approach, which uses atomic norm minimization to manifest the signal sparsity in the continuous parameter domain, has been proposed for several signal processing applications \cite{Atom1}. Under certain conditions, atomic norm minimization can achieve exact sparse signals reconstruction,
avoiding the effects of basis mismatch which can plague grid-based CS techniques.

In this paper, we propose super-resolution mmWave channel estimators based on atomic norm minimization
for both the HB system and GSM system. First, the pilot-assisted channel estimator is formulated as a convex optimization problem under the atomic norm minimization framework that exploits the sparsity in the continuous AoD/AoA domains. Then to account for the slowly time-varying nature of the block-fading channels, the data-aided channel estimator is formulated as a combined atomic norm and $\ell_1$-norm minimization problem, which is non-convex. Here the $\ell_1$-norm is to exploit the sparsity in the demodulation errors. Moreover, we develop computationally efficient non-convex methods to solve both channel estimation formulations based on non-convex factorization and conjugate gradient descent (CGD). Extensive simulation results are provided to illustrate the superior performance of the proposed new channel estimators compared with the existing CS-based methods.

The remainder of this paper is organized as follows. Section \uppercase\expandafter{\romannumeral2}
describes the mmWave channel model and the signal models for both the HB system and GSM system.
Section \uppercase\expandafter{\romannumeral3} gives the formulations of both pilot-assisted and
data-aided channel estimators. Section \uppercase\expandafter{\romannumeral4} presents the proposed
non-convex method for solving both channel estimation formulations. Simulation results are given
in Section \uppercase\expandafter{\romannumeral5}. Finally, Section VI concludes the paper.
%
%Throughout this paper, the following notations are used for description. $\mathbf{A}$, $\mathbf{a}$ and a denote the matrix, vector and scalar, respectively. $\mathbf{A}^*$, $\mathbf{A}^T$, $\mathbf{A}^H$ and $\mathbf{A}^{\dag}$ represent the matrix conjugate, transpose, conjugate transpose, and pseudo-inverse, respectively. diag$(\mathbf{a})$ represents a diagonal matrix with the entries of $\mathbf{a}$ on its diagonal.
%$\mathbf{A}(n)$ denotes the $n$-th column of $\mathbf{A}$, a(n) denotes the $n$-$th$ entry of $\mathbf{a}$. $\mathrm{vec}(\mathbf{A})$ is a vector obtained through the vectorization of $\mathbf{A}$.
%For $M \times N$ matrices $\mathbf{A}$ and $\mathbf{B}$, $\mathbf{A} \otimes \mathbf{B}$ denotes the
%$M^2 \times N^2$ matrix of Kronecker
%product between $\mathbf{A}$ and $\mathbf{B}$, and $\mathbf{A}$ $\odot$ $\mathbf{B}$ denotes the
%$M^2\times N$ matrix of Khatri-Rao product defined as $\mathbf{A}$ $\odot$ $\mathbf{B} = [\mathbf{A}(1)$
%$\otimes$ $\mathbf{B}(1),...,\mathbf{A}(N)$ $\otimes$ $\mathbf{B}(N)]$. We use $\mathbb{C}$ and $\mathbb{R}$ to denote the field of complex and real numbers, respectively. $\mathbb{E}[\cdot]$ is the
%expectation operator, $\mathbf{I}_n$ denotes a $n\times n$ identity matrix, $\mathbf{O}_{m\times n}$ denotes a $m\times n$ null matrix, $\mathbf{0}_n$ denotes a $n\times 1$ null vector,
%$m \choose n$ denotes the binomial coefficient and $\lfloor \cdot \rfloor$ denotes the floor function.

\section{System Descriptions}%1级小标题1

In this section, we first present the mmWave channel model and then the signal models for the HB system and GSM system, respectively.

\subsection{Channel Model}

We consider a wireless communication system operating at mmWave band. The transmitter has $N_t$ antennas and $n_t$ RF chains, and the receiver has $N_r$ antennas and $n_r$ RF chains, where max$(n_t, n_r)\leq$ min$(N_t,N_r)$. Both the transmit and receive antennas are uniform linear arrays.

We assume a geometric MIMO channel model \cite{Channel-model}
that has $L$ scatterers during a time block. Thus, the channel matrix in a time block can be expressed as
\begin{align}\label{Chan}
\mathbf{H}=\sum\limits_{l=1}^L {\alpha_l}\mathbf{a}_R(\phi_l)\mathbf{a}_T^H(\theta_l) \in \mathbb{C}^{N_r \times N_t},
\end{align}
where $\alpha_l \sim \mathcal{CN}(0,\sigma_l^2)$ is the complex gain of the $l$-th path, $\sigma_l^2$ is the average power gain of the $l$-th path.
Assuming that the antenna arrays are installed in the horizontal
direction, we denote $\phi_l = \sin(\bar{\phi}_l)\in [0,1)$ and $\theta_l = \sin(\bar{\theta}_l)\in [0,1)$ as the departure and arrival directions of the $l$-th path, respectively, where $\bar{\phi}_l$ and $\bar{\theta}_l$ are the physical azimuth angles of departure and arrival (AoD/AoA), respectively. ${\mathbf{a}_T}(\theta_l)$ represents the normalized transmit array response vector at the direction of $\theta_l$ given by
\begin{equation}
\mathbf{a}_T(\theta_l) = \frac{1}{\sqrt{N_t}} {\left[{1,{e^{j\frac{2\pi }{\lambda}d\theta_l}},...,{e^{j(N_t-1)\frac{{2\pi}}{\lambda}d\theta_l}}}\right]^T} \in \mathbb{C}^{N_t \times 1},
\end{equation}
where $\lambda$ is the wavelength, $d$ is the inter-antenna element spacing with $d \geq \lambda/2$. The antenna array response $\mathbf{a}_R(\phi_l)$ at the receiver can be written similarly.

Thus, (1) can be expressed compactly as
\begin{equation}
\mathbf{H} = \mathbf{A}_R \mathbf{\Lambda} \mathbf{A}_T^H \in \mathbb{C}^{N_r \times N_t},
\end{equation}
where $\mathbf{\Lambda}= \mathrm{diag}([\alpha_1,\alpha_2,...,\alpha_L]^T)$, and
\begin{eqnarray}
\mathbf{A}_T &=&[\mathbf{a}_T(\theta_1),...,\mathbf{a}_T(\theta_L)] \in \mathbb{C}^{N_t \times L},\\
\mathbf{A}_R &=&[\mathbf{a}_R(\phi_1),...,\mathbf{a}_R(\phi_L)] \in \mathbb{C}^{N_r \times L}.
\end{eqnarray}

\subsection{Signal Models}

In this paper, we will consider both the HB system and GSM system illustrated in Fig. 1(a) and Fig. 1(b), respectively.
\begin{figure}[!htb]
\begin{center}
\subfigure[The HB system.]{
%\label{fig:subfig:a} %% 第一幅图的标签
\includegraphics[scale=0.49]{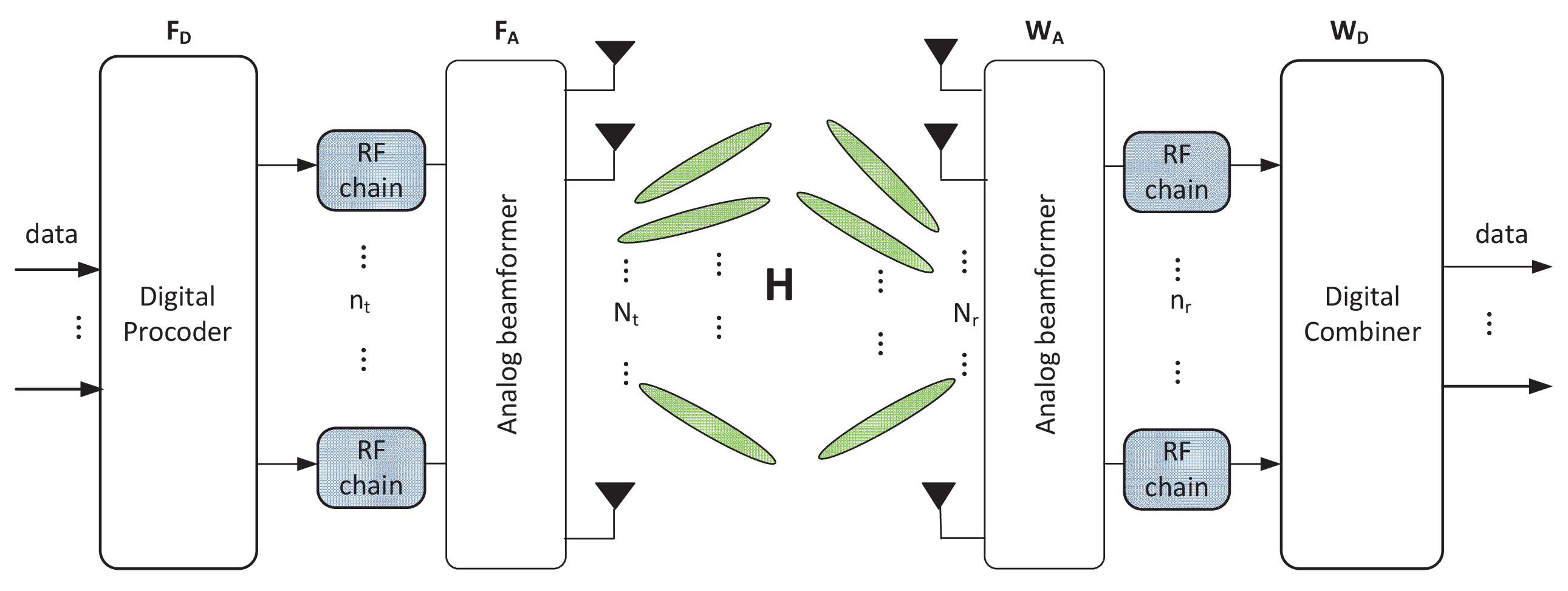}}
%\hspace{10in}
\subfigure[The GSM system.]{
%\label{fig:subfig:b} %% 第二幅图的标签
\includegraphics[scale=0.49]{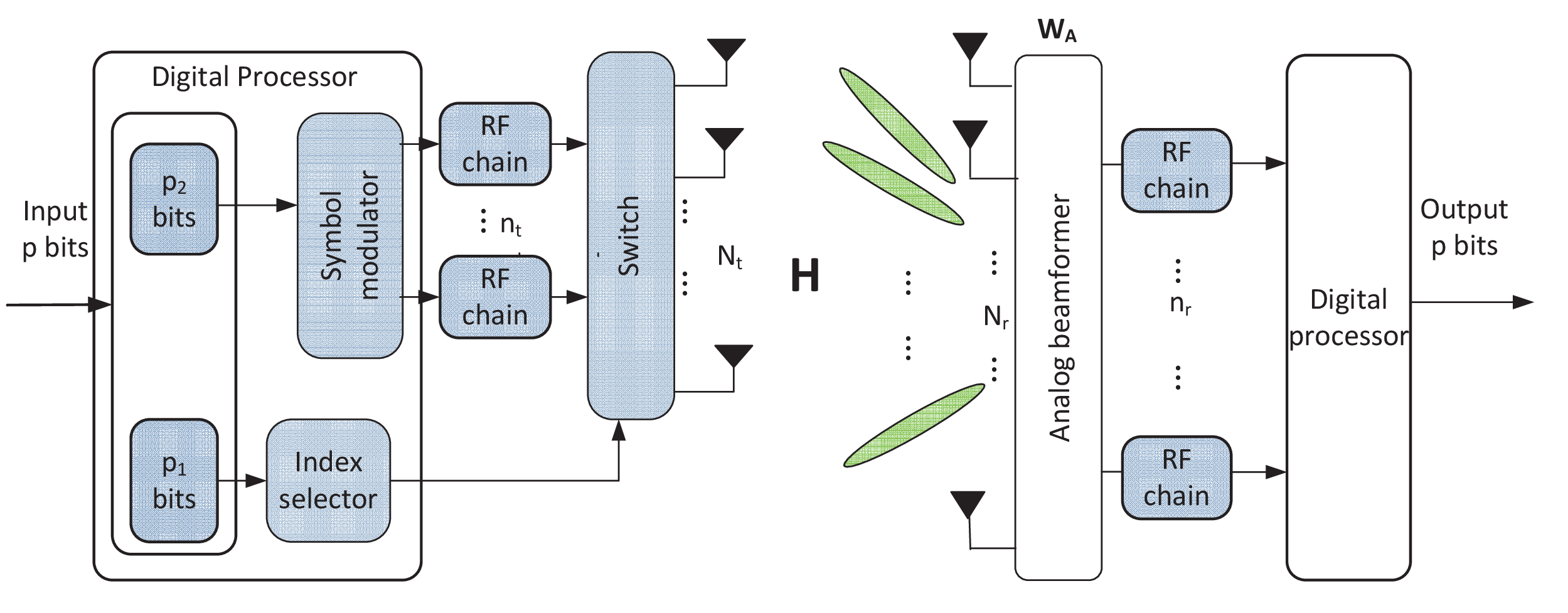}}
%\captionsetup{font=small}
\caption{The HB system and GSM system.}
%\label{fig:subfig} %% label for entire figure
\end{center}
\end{figure}
For both systems, the transmitted signal during the $k$-th time slot is denoted by $\mathbf{x}_k \in \mathbb{C}^{N_t\times 1}, k=1,2,...,K$. The $N_r$-dimensional signal at the receive antenna array is processed by a linear filter $\mathbf W \in \mathbb{C}^{N_r \times n_r}$, resulting in the following signal at the output of the receive RF chain
\begin{equation}
\mathbf{y}_k = \mathbf{W}^H \mathbf{H} \mathbf{x}_k + \mathbf{W}^H \mathbf{n}_k \in \mathbb{C}^{{n_r}
\times 1},
\end{equation}
where $\mathbf{n}_k \sim \mathcal{CN}(0,\sigma^2 \mathbf{I}_{N_r})\in\mathbb{C}^{N_r\times 1}$ is the additive white Gaussian noise and $\mathbf{I}_{N_r}$ denotes a $N_r \times N_r$ identity matrix. Denoting $\mathbf{X} = [\mathbf{x}_1,\mathbf{x}_2,...,\mathbf{x}_K]$, $\mathbf{Y} = [\mathbf{y}_1,\mathbf{y}_2,...,\mathbf{y}_K]$ and
$\mathbf{Q} = \mathbf{W}^H[\mathbf{n}_1,\mathbf{n}_2,...,\mathbf{n}_K]$, then (5) can be written as
\begin{equation}\label{Rece}
\mathbf{Y} = \mathbf{W}^H \mathbf{H} \mathbf{X} + \mathbf{Q}.
\end{equation}

\subsubsection{HB System}

In the HB system, the transmitted signal is given by $\mathbf{x}_k = \mathbf{F}\mathbf{s}_k = \mathbf{F}_A \mathbf{F}_D \mathbf{s}_k$, where $\mathbf{s}_k \in {\cal M}^{M_s}$ is the data symbol vector, with ${\cal M}$ being the symbol constellation set, $M_s \leq n_t$ is the number of data streams, $\mathbf{F}_D \in \mathbb{C}^{n_t \times M_s}$ is the digital precoder that adjusts both amplitudes and phases, and $\mathbf{F}_A \in \mathbb{C}^{N_t \times n_t}$ is the analog RF precoder that only adjusts phases. At the receiver, the received signal is firstly passed through the analog RF combiner $\mathbf{W}_A \in \mathbb{C}^{N_r \times n_r}$ and then the baseband combiner $\mathbf{W}_D \in \mathbb{C}^{n_r \times M_s}$. Hence the combiner $\mathbf{W}$ in (6) can be written as $\mathbf{W} = \mathbf{W}_A \mathbf{W}_D$. Note that all elements of $\mathbf{F}_A$ and $\mathbf{W}_A$ should have constant amplitudes. Examples of the analog filters $\mathbf{F}_A$, $\mathbf{W}_A$, and digital filters $\mathbf{F}_D$, $\mathbf{W}_D$ can be found in \cite{Beamforming}.

In particular, during the pilot training stage, we set $M_s = n_t = n_r$ and the digital filters $\mathbf{F}_D=\mathbf{I}_{n_t}$,  $\mathbf{W}_D = \mathbf{I}_{n_r}$. The DFT beamformers can be employed as the analog filters, given by
\begin{eqnarray}\label{DFT}
\mathbf{F}_A &=& [\mathbf{a}_T(\bar{\theta}_1),...,\mathbf{a}_T(\bar{\theta}_{n_t})] \in \mathbb{C}^{N_t \times n_t},\\
\mathbf{W}_A &=& [\mathbf{a}_R(\bar{\phi}_1),...,\mathbf{a}_R(\bar{\phi}_{n_r})] \in \mathbb{C}^{N_r \times n_r},
\end{eqnarray}
where
\begin{eqnarray}
\bar{\theta}_i &=& \bar{\theta}_1 + \frac{2}{N_t}(i-1),~~i=1,2,...,n_t,\\
\bar{\phi}_j &=& \bar{\phi}_1 + \frac{2}{N_r}(j-1),j=1,2,...,n_r,
\end{eqnarray}
with $\bar{\theta}_i$, $\bar{\phi}_j \in [0,1)$ denoting the pointing directions of the $i$-th transmit beam and the $j$-th receive beam, respectively.

\subsubsection{GSM System}

In the GSM system, in each time slot, only $n_t$ out of $N_t$ transmit antennas are activated to transmit data while the other $N_t-n_t$ transmit antennas remain idle. The information bits are conveyed by not only the modulation symbols but also the indices of the active antennas. As shown in Fig. 1(b), at the transmitter, a block of $p = p_1+ p_2$ information bits is divided into two parts. The first $p_1 = \lfloor \log_2 {N_t \choose n_t} \rfloor$ bits are fed to the index selector to determine the indices of $n_t$ active antennas $\cal{U}$$= \{u_1,u_2,...,u_{n_t}\}$, where $u_m \in \{1,2,...,N_t\}$ for $m=1,2,...,n_t$ and $u_1<u_2<...<u_{n_t}$. Note that among $N_t \choose n_t$ possible transmit antenna combinations, only $2^{p_1}$ transmit antenna combinations are permitted and the other $N_t \choose n_t -2^{p_1}$ combinations are illegal. The remaining $p_2 = n_t \log_2 M$ bits are then fed to the symbol modulator to generate $n_t$ modulation symbols each drawn from a constellation alphabet $\cal M$ of cardinality $M$ and carried on an antenna indexed by an element in $\cal U$, resulting in the transmitted signal $\mathbf{x}_k =[x_k(1),x_k(2),...,x_k(N_t)]^T$ in (6)-(7), where
\begin{align}
{{x}_k}(n) = \left\{ \begin{gathered}
{s_m} \in {\cal M},~n = {u_m} \in \cal U, \hfill \\
0,~~~~~~~~~~ n  \notin \cal U, \hfill
\end{gathered}  \right.\quad n=1,2,\ldots,N_t.
\end{align}
Hence $x_k(n) \in {{\cal M}\cup\{0\}}$ and $\|\mathbf{x}_k\|_0 = n_t$. The mapping of $p_1$ bits for index selection can be implemented by using a look-up table or the combinatorial method \cite{GSMMIMO}. In the signal model (6)-(7), for the GSM system, we have $\mathbf{W}=\mathbf{W}_A \in {\mathbb C}^{N_r \times n_r}$. For example, the DFT beamformers in (8)-(9) can be employed.

\section{Channel Estimation Based on Atomic Norm Minimization}

In this section, we formulate the mmWave channel estimation problem as an atomic norm based sparse recovery problem in the continuous AoD/AoA domains. Both the pilot-assisted and data-aided channel estimators are developed.

From (\ref{Chan}), estimating the channel matrix $\mathbf{H}$ is equivalent to estimating the parameters of the $L$ paths, i.e., $\{\alpha_l, \phi_l, \theta_l\}_{l=1}^L.$ Since the number of scatters $L$ in the mmWave channel is typically small, the system exhibits sparsity that can be exploited for channel estimation purpose. To begin with, we vectorize $\mathbf{Y}$ in (7) to obtain
\begin{equation}\label{Vect}
%\begin{split}
\tilde{\mathbf{y}} = \mathrm{vec}(\mathbf{Y}) = (\mathbf{X}^T \otimes \mathbf{W}^H)\mathbf{\tilde{h}} + \mathbf{\tilde{q}}
= (\mathbf{X}^T \otimes \mathbf{W}^H) (\mathbf{A}^\ast_T \odot\mathbf{A}_R) \bm{\alpha} + \mathbf{\tilde{q}},
%\end{split}
\end{equation}
where $\bm{\alpha}=[\alpha_1,\alpha_2,...,\alpha_L]^T$, $\mathbf{X}^T \otimes \mathbf{W}^H \in \mathbb{C}^{n_r K\times {N_tN_r}}$ with $\otimes$ being the Kronecker product, $\tilde{\mathbf{h}} = \mathrm{vec}(\mathbf{H})\in \mathbb{C}^{N_tN_r\times 1}$ and $\mathbf{\tilde{q}} = \mathrm{vec}(\mathbf{Q}) \in \mathbb{C}^{N_tN_r\times 1}$. $\mathbf{A}^\ast_T \odot \mathbf{A}_R$ is an $N_t N_r \times L$ matrix in which each column has the form $\mathbf{a}^\ast_T(\theta_l)\otimes \mathbf{a}_R(\phi_l)$, with $\ast$ being the conjugation operation and $\odot$ being the Khatri-Rao product.

\subsection{Channel Estimation Based on On-grid CS Algorithm}

Before describing our proposed mmWave channel estimators, we briefly discuss some existing CS-based
mmWave channel estimation methods \cite{OMP}.
%\subsubsection{CS-L1 based mmWave channel estimators:}
Estimating $\mathbf{H}$ is equivalent to jointly estimating the unknown parameters $\bm{\alpha}$, $\mathbf{A}_T$ and $\mathbf{A}_R$ from the noisy observations $\tilde{\mathbf{y}}$ in (\ref{Vect}), which is a non-linear problem. However, it can be linearized by using an overcomplete dictionary matrix defined as $\mathbf{\tilde{A}}_T^\ast \odot \mathbf{\tilde{A}}_R$ \cite{L1} with
\begin{eqnarray}
\mathbf{\tilde{A}}_T &=&[\mathbf{a}_T(\tilde{\theta}_1'),\mathbf{a}_T(\tilde{\theta}_2'),...,\mathbf{a}_T(\tilde{\theta}_{\tilde{J}}')] \in \mathbb{C}^{N_t \times \tilde{J}},\\
\mathbf{\tilde{A}}_R &=&[\mathbf{a}_R(\tilde{\phi}_1'),\mathbf{a}_R(\tilde{\phi}_2'),...,\mathbf{a}_R(\tilde{\phi}_{\tilde{J}}')] \in \mathbb{C}^{N_r \times \tilde{J}},
\end{eqnarray}
where $\{\tilde{\theta}_{j}'\}_{j=1}^{\tilde{J}}$ and $\{\tilde{\phi}_{j}'\}_{j=1}^{\tilde{J}}$ denote sets of uniformly spaced points in the interval $[0,1)$, and $\tilde{J}$ is the number of columns of $\mathbf{\tilde{A}}_T$ or $\mathbf{\tilde{A}}_R$ where $\tilde{J} \gg L$. For sufficiently large $\tilde{J}$, the angles are densely sampled. Let $\tilde{\bm{\alpha}}=[\tilde{\alpha}_1',\tilde{\alpha}_2',...,\tilde{\alpha}_{\tilde{J}}']^T \in \mathbb{C}^{\tilde{J}\times 1}$ be the sparse vector whose non-zero elements correspond to $\bm{\alpha}$ in (\ref{Vect}). Thus, the non-linear parameter estimation problem is reduced to the following problem \cite{OMPL0}:
\begin{equation}\label{L0}
\begin{aligned}
\hat{\bm{\alpha}}' = &\arg \min \limits_{\tilde{\bm{\alpha}}\in \mathbb{C}^{\tilde{J}\times 1}}
\left\| {\tilde{\bm{\alpha}}} \right\|_0\\
\rm{s.t.}~~~&\left\| {{{\tilde{\mathbf{y}}}} - (\mathbf{X}^T \otimes \mathbf{W}^H) (\tilde{\mathbf{A}}^\ast_T \odot\tilde{\mathbf{A}}_R) \tilde{\bm{\alpha}}} \right\|_2^2 < \epsilon.
\end{aligned}
\end{equation}
%\begin{equation}\label{L0}
%\begin{flalign}
%\hat{\bm{\alpha}}' &= \arg \min \limits_{\tilde{\bm{\alpha}}\in \mathbb{C}^{\tilde{J}\times 1}}
%\frac{1}{2}\left\| {{{\tilde{\mathbf{y}}}} - (\mathbf{X}^T \otimes \mathbf{W}^H) (\tilde{\mathbf{A}}^\ast_T \odot\tilde{\mathbf{A}}_R) \tilde{\bm{\alpha}}} \right\|_2^2\\
%\rm{s.t.}~~&\left\| {\tilde{\bm{\alpha}}} \right\|_0 = L,
%\end{flalign}
%\end{equation}
Note that (\ref{L0}) is non-convex. In practice, greedy algorithms such as orthogonal matching pursuit (OMP) \cite{OMP} can be used to find a suboptimal solution to (\ref{L0}).

Alternatively, the $\ell_1$-norm regularization, i.e., $\|\tilde{\bm{\alpha}}\|_1 = \sum_{j=1}^{\tilde{J}}|\tilde{\alpha}_j|$, can be employed and the optimization problem can be written as:
\begin{equation}\label{L1}
\hat{\bm{\alpha}}' = \arg \min \limits_{\tilde{\bm{\alpha}}\in \mathbb{C}^{\tilde{J}\times 1}}
\frac{1}{2}\left\| {{{\tilde{\mathbf{y}}}} - (\mathbf{X}^T \otimes \mathbf{W}^H) (\tilde{\mathbf{A}}^\ast_T \odot\tilde{\mathbf{A}}_R) \tilde{\bm{\alpha}}} \right\|_2^2 + \mu {\left\| {\tilde{\bm{\alpha}}} \right\|_1},
\end{equation}
where $\mu > 0$ is the weight factor. As (\ref{L1}) is convex, it can be solved with standard convex solvers. In this paper, we name the algorithm that solves (\ref{L1}) the CS-L1 algorithm.

Both the OMP and CS-L1 algorithms can super-resolve the angles of the sparse signal under certain conditions on the dictionary matrix $\mathbf{\tilde{A}}_T^\ast \odot \mathbf{\tilde{A}}_R$. The estimated channel is then given by
\begin{equation}\label{L1H}
\hat{\mathbf{H}} = \mathbf{\tilde{A}}_R {\rm diag}(\hat {\bm{\alpha}}') \mathbf{\tilde{A}}_T^H.
\end{equation}
However, the angles of interest are discretized into a number of grids, and the actual angles may not exactly reside on the grid points. Such an off-grid problem can deteriorate the channel estimation performance.

\subsection{Sparsity Enforcement Via Atomic Norm Minimization}
%\vspace{-1.6ex}
To solve the off-grid problem, we employ the 2D atomic norm to enforce the sparsity of $\tilde{\mathbf{h}}$. First, we briefly introduce the concept of 2D atomic norm \cite{Norm}.
Suppose that $\mathbf{c}(\theta,\phi)$ is the building block (called 2D atom) of a class of signals. The 2D atomic set is defined as $\mathcal{A}=\{\mathbf{c}(\theta,\phi)|\theta \in [0,1),\phi \in [0,1)\}$.
%\begin{equation}
%\mathcal{A}=\{\mathbf{c}(\theta,\phi)|\theta \in [0,1),\phi \in [0,1)\}.
%\end{equation}

Then the 2D atomic norm of any signal $\mathbf{p}$ in the mentioned signal class with respect to $\mathcal{A}$ is defined as
\begin{equation}\label{Atomnorm}
\begin{split}
\|\mathbf{p}\|_{\mathcal{A}} &= \inf\{\gamma > 0: \mathbf{p} \in \gamma \mathrm{conv}(\mathcal{A})\}\\
&=\inf\limits_{\theta_j \in [0,1),\phi_j \in [0,1),\alpha_j\in \mathbb{C}} \biggl\{\sum_{j=1}^{J} |\alpha_j|: \mathbf{p}= \sum_{j=1}^{J} \alpha_j \mathbf{c}(\theta_j,\phi_j)\biggr\},
\end{split}
\end{equation}
where $\inf\{\cdot\}$ denotes the infimum of the input set, and $\mathrm{conv}(\mathcal{A})$ denotes the convex hull of $\mathcal{A}$. From (\ref{Vect}), the class of signals is $\tilde{\mathbf{h}}=(\mathbf{A}^\ast_T \odot\mathbf{A}_R)\bm{\alpha}=\sum\limits_{l=1}^L{\alpha_l}\mathbf{c}(\theta_l,\phi_l)$.
Therefore the atom is of the form $\mathbf{c}(\theta_l,\phi_l)=\mathbf{a}^\ast_T(\theta_l)\otimes \mathbf{a}_R(\phi_l) \in \mathbb{C}^{N_t N_r \times 1}$. The 2D atomic norm for $\tilde{\mathbf{h}}$ is then
\begin{equation}
\begin{array}{l}
{\left\| {\tilde{\mathbf{h}}} \right\|_{\mathcal{A}}} = \mathop {\inf }\limits_{\scriptstyle{\mathbf{c}(\theta_l,\phi_l)} \in \mathcal{A} \hfill\atop
\scriptstyle{\alpha_l} \in \mathbb{C}}\left\{\sum\limits_{l=1}^L |{\alpha_l}|:\tilde {\mathbf{h}} = \sum\limits_{l=1}^L {{\alpha_l}\mathbf{c}(\theta_l,\phi_l)} \right\}.
\end{array}
\end{equation}

On this basis, an optimization problem for channel estimation will be formulated using the following equivalent form of the 2D atomic norm \cite{SDP}:
\begin{equation}\label{SDP}
\begin{aligned}
{\left\|{\tilde{\mathbf{h}}} \right\|_{\mathcal{A}}} &= \mathop{\inf}\limits_{\mathbf{V}\in \mathbb{C}^{(2N_t-1)\times (2N_r-1)}\hfill\atop \varepsilon \in \mathbb{R}}
\frac{1}{2 N_t N_r}\mathrm{Tr}({\mathcal T}(\mathbf{V})) + \frac{\varepsilon}{2}\\
\rm{s.t.}~~\mathbf{\Psi}~~&=\left[{\begin{array}{*{20}{c}}
{{\mathcal T}(\mathbf{V})}&{\tilde{\mathbf{h}}}\\
{{{\tilde{\mathbf{h}}}^H}}&\varepsilon
\end{array}} \right] \succeq 0,
\end{aligned}
\end{equation}
where $\mathrm{Tr}(\cdot)$ is the trace operator and $\mathbf{V}\in \mathbb{C}^{(2N_t-1)\times (2N_r-1)}$ is defined as
\begin{equation}
\mathbf{V} = [\mathbf{v}_{-N_r+1},\mathbf{v}_{-N_r+2},...,\mathbf{v}_{N_r-1}]
\end{equation}
with $\mathbf{v}_{g} = [v_g(-N_t+1),v_g(-N_t+2),...,v_g(N_t-1)]^T \in \mathbb{C}^{(2N_t-1)\times 1},
g=-N_r+1,-N_r+2,...,N_r-1$.
%$\left[{\begin{array}{*{20}{c}}
%{{\mathcal T}(\mathbf{V})}&{\tilde{\mathbf{h}}}\\
%{{{\tilde{\mathbf{h}}}^H}}&\varepsilon
%\end{array}} \right] \in \mathbb{C}^{(N_tN_r+1)\times(N_tN_r+1)}$ and
$\mathcal{T}(\mathbf{V})$ is a block Toeplitz matrix defined as
\begin{equation}\label{Toeplitz}
\begin{array}{l}
{\mathcal T}(\mathbf{V}) = \left[ {\begin{array}{*{20}{c}}
{\mathrm{Toep}({\mathbf{v}_0})}&{\mathrm{Toep}({\mathbf{v}_{-1}})}&{...}&{\mathrm{Toep}({\mathbf{v}_{-N_r+1}})}\\
{\mathrm{Toep}({\mathbf{v}_1})}&{\mathrm{Toep}({\mathbf{v}_0})}&{...}&{\mathrm{Toep}({\mathbf{v}_{-N_r+2}})}\\
{\begin{array}{*{20}{c}}
 \vdots
\end{array}}& \vdots & \ddots & \vdots \\
{\mathrm{Toep}({\mathbf{v}_{N_r-1}})}&{\mathrm{Toep}({\mathbf{v}_{N_r-2}})}&{...}&{\mathrm{Toep}({\mathbf{v}_0})}
\end{array}} \right] \in \mathbb{C}^{N_t N_r\times N_t N_r},
\end{array}
\end{equation}
where $\mathrm{Toep}(\cdot)$ denotes the Toeplitz matrix whose first column is the last $N_t$ elements of the input vector. More specifically, we have
\begin{equation}
\begin{array}{l}
\mathrm{Toep}(\mathbf{v}_g) = \left[{\begin{array}{*{20}{c}}
{{\mathrm{v}_g}(0)}&{{\mathrm{v}_g}(-1)}&{...}&{{\mathrm{v}_g}(-N_t+1)}\\
{{\mathrm{v}_g}(1)}&{{\mathrm{v}_g}(0)}&{...}&{{\mathrm{v}_g}(-N_t+2)}\\
{\begin{array}{*{20}{c}}
 \vdots
\end{array}}& \vdots & \ddots & \vdots \\
{{\mathrm{v}_g}(N_t-1)}&{{\mathrm{v}_g}(N_t-2)}&{...}&{{\mathrm{v}_g}(0)}
\end{array}} \right] \in \mathbb{C}^{N_t \times N_t},\\~~~~~~~~~~g =-N_r+1,-N_r+2,...,N_r-1.
\end{array}
\end{equation}
%In the following, we will introduce a proposition to simplify the channel estimators to be proposed in Section \uppercase\expandafter{\romannumeral3} by using the format of (\ref{SDP}) and the spatial sparsity of mmWave channels. Based on the fact that $L$ is usually small, we show in the following proposition that under some conditions the rank of $\mathbf{\Psi}$ is smaller than the upper bound on the number of paths, e.g., $\bar L$. The proof is given in Appendix A.

\subsection{Pilot-assisted Channel Estimator}

Assuming that $\mathbf{X}$ in (\ref{Vect}) contains known pilot symbols either for the HB system or the GSM system, we can formulate the following optimization problem for the pilot-assisted channel estimator:
\begin{equation}\label{Pilo}
\begin{array}{l}
{\hat{\mathbf{h}}} = \arg \min \limits_{\tilde{\mathbf{h}}\in \mathbb{C}^{N_tN_r\times 1}}
\frac{1}{2}\left\| {{{\tilde{\mathbf{y}}}} - (\mathbf{X}^T \otimes \mathbf{W}^H){{\tilde{\mathbf{h}}}}} \right\|_2^2 + \mu {\left\| {{{\tilde{\mathbf{h}}}}} \right\|_\mathcal{A}}.
\end{array}
\end{equation}
Applying (\ref{SDP}), then (\ref{Pilo}) can be transformed to the following semidefinite program (SDP):
\begin{equation}\label{PSDP}
\begin{aligned}
{\hat{\mathbf{h}}} = & \arg \min \limits_{\tilde{\mathbf{h}}\in \mathbb{C}^{N_tN_r\times 1}, \varepsilon\in \mathbb{R}\hfill\atop \mathbf{V}\in \mathbb{C}^{(2N_t-1)\times(2N_r-1)}} \frac{1}{2}\left\| {{{\tilde{\mathbf{y}}}} - (\mathbf{X}^T \otimes \mathbf{W}^H){{\tilde{\mathbf{h}}}}} \right\|_2^2 + \frac{\mu}{2N_tN_r}\mathrm{Tr}({\mathcal T}({\mathbf{V}})) + \frac{\mu \varepsilon}{2}\\
\rm{s.t.}~~&\left[{\begin{array}{*{20}{c}}
{{\mathcal T}({\mathbf{V}})}&{\tilde{\mathbf{h}}}\\
{{\tilde{\mathbf{h}}}^H}&{\varepsilon}
\end{array}} \right] \succeq 0.
\end{aligned}
\end{equation}
The above problem is convex, so it can be solved efficiently using a convex solver. We denote the solution to (\ref{PSDP}) as $\hat{\mathbf{h}}$. The estimate of the channel matrix is then $\hat{\bold{H}}={\mathrm{vec}}^{-1}(\hat{\mathbf{h}})$. Note that the number of paths $L$ is not needed in the above formulationn. We name the channel estimator given by (\ref{PSDP}) as the pilot-assisted estimator based on 2D atomic norm (Atom-pilot).

\subsection{Data-aided Channel Estimator}

We now consider a total of $T+1$ transmission blocks of the form of (7), i.e.,
\begin{equation}\label{Re28}
\mathbf{Y}_t = \mathbf{W}_t^H \mathbf{H}_t \mathbf{X}_t + \mathbf{Q}_t, ~t=0,1,...,T,
\end{equation}
where $t=0$ corresponds to the pilot block, i.e., $\mathbf{X}_0$ contains known pilot symbols and all other blocks, i.e., $\mathbf{X}_t$ for $t=1,2,...,T$ are data blocks. Traditionally, it is assumed that the channel remains invariant during the $T+1$ blocks, i.e., $\mathbf{H}_0=\mathbf{H}_1=...=\mathbf{H}_T$ and the estimated channel $\hat{\mathbf{H}}_0$ during the pilot block is used to demodulate the data symbols
($\mathbf{S}_t$ for HB system and
$\mathbf{X}_t$ for GSM system) during all data blocks $t=1,2,...,T$. However, in practice, the channel may be slowly varying across different time blocks, i.e., $\mathbf{H}_t=\mathbf{H}_{t-1}+\Delta\mathbf{H}_t$, $t=1,2,...,T$. Here we consider a data-aided channel estimation scheme, where at $t=0$ pilot symbols are used to estimate $\mathbf{H}_0$. In the subsequent data blocks, $t=1,2,...,T$, first the previous channel estimate $\hat{\mathbf{H}}_{t-1}$ is used to demodulate the data in the current block; then the demodulated data symbols in the current block are employed to obtain the current channel estimate $\hat{\mathbf{H}}_t$. Next we describe the corresponding formulations for the HB and GSM systems, respectively.

\subsubsection{HB System}

For the HB system, we first perform an initial estimate of $\mathbf{X}_t$ using the channel estimate $\hat{\mathbf{H}}_{t-1}$ from the previous time block, $t=1,2,...,T$. Recall that $\mathbf{X}_t = \mathbf{F}_t \mathbf{S}_t$ where the transmit beamformer $\mathbf{F}_t$ is formed based on the channel estimate $\hat{\mathbf{H}}_{t-1}$. We then demodulate the data symbols $\mathbf{S}_t$ by solving
\begin{equation}\label{Dete}
\begin{split}
\check{\mathbf{S}}_t =& \arg \min_{\mathbf{S}_t\in {\cal M}^{n_t \times K}} \| \mathbf{Y}_t - \mathbf{W}_t^H \hat{\mathbf{H}}_{t-1} \mathbf{F}_t \mathbf{S}_t\|_F
\end{split}
\end{equation}
either optimally or suboptimally. Define the data symbol error matrix as
%\begin{equation}
$\mathbf{E}_t = \mathbf{S}_t - \check{\mathbf{S}}_t.$
%\end{equation}
Then (\ref{Re28}) can be written as
\begin{equation}\label{DataHB}
\mathbf{Y}_t = \mathbf{W}_t^H \mathbf{H}_t \mathbf{F}_t (\check{\mathbf{S}}_t + \mathbf{E}_t) + \mathbf{Q}_t.
\end{equation}
Note that the combiner $\mathbf{W}_t$ in (\ref{Dete}) and (\ref{DataHB}) is also formed based on $\hat{\mathbf{H}}_{t-1}$. Under the normal system operating condition, the demodulation error rate should be low; that is, the error matrix $\mathbf{E}_t$ is sparse. Thus, to estimate $\mathbf{H}_t$ and $\mathbf{E}_t$ from (\ref{DataHB}), we formulate the following optimization problem, where for notational simplicity we drop the subscript $t$:
\begin{equation}\label{DATAHB}
%\begin{array}{l}
(\hat{\mathbf{h}},\hat{\mathbf{e}}) = \arg \min \limits_{\tilde{\mathbf{h}}\in \mathbb{C}^{N_tN_r\times 1}\hfill\atop\mathbf{e}\in \mathbb{C}^{KM_s\times 1}}
\frac{1}{2}\left\|{{{\tilde{\mathbf{y}}}} - \left((\check{\mathbf{S}}^T+[\mathrm{vec}^{-1}(\mathbf{e})]^T)\mathbf{F}^T \otimes \mathbf{W}^H \right) {{\tilde{\mathbf{h}}}}}\right\|_2^2 + \mu{\left\|{{{\tilde{\mathbf{h}}}}}\right\|_\mathcal{A}} + \lambda\|\mathbf{e}\|_1,
%\end{array}
\end{equation}
where $\mathbf{e} = \mathrm{vec}(\mathbf{E})\in \mathbb{C}^{KM_s\times 1}$, $\lambda > 0$ is the weight factor. Substituting (\ref{SDP}) to (\ref{DATAHB}), we obtain the following constrained optimization problem:
\begin{equation}\label{DSDPHB}
\begin{aligned}
(\hat{\mathbf{h}},\hat{\mathbf{e}}) = & \arg \min \limits_{\tilde{\mathbf{h}}\in \mathbb{C}^{N_tN_r\times 1},\mathbf{e}\in \mathbb{C}^{KM_s\times 1}\hfill\atop \mathbf{V}\in \mathbb{C}^{(2N_t-1)\times(2N_r-1)},\varepsilon\in \mathbb{R}}
\frac{1}{2}\left\| {{{\tilde{\mathbf{y}}}} - ((\check{\mathbf{S}}^T+[\mathrm{vec}^{-1}(\mathbf{e})]^T)\mathbf{F}^T \otimes \mathbf{W}^H){{\tilde{\mathbf{h}}}}} \right\|_2^2 + \frac{\mu}{2N_tN_r}\mathrm{Tr}({\mathcal T}({\mathbf{V}})) \\&+ \frac{\mu \varepsilon}{2} + \lambda \|\mathbf{e}\|_1\\
\rm{s.t.}~~~~&\mathbf \Psi=\left[{\begin{array}{*{20}{c}}
{{\mathcal T}({\mathbf{V}})}&{\tilde{\mathbf{h}}}\\
{{\tilde{\mathbf{h}}}^H}&{\varepsilon}
\end{array}} \right] \succeq 0.
\end{aligned}
\end{equation}
Note that the above problem is non-convex due to the product term of $\tilde{\mathbf{h}}$ and $\mathbf{e}$. We will propose an efficient method to solve (\ref{DSDPHB}) in the next section. Given the solution $\hat{\mathbf{h}}$ and $\hat{\mathbf{e}}$ to (\ref{DSDPHB}), we can obtain the estimates of the channel matrix $\hat{\mathbf{H}}={\mathrm{vec}}^{-1}(\hat{\mathbf{h}})$ and the sparse error matrix $\hat{\mathbf{E}}={\mathrm{vec}}^{-1}(\hat{\mathbf{e}})$. Finally we can refine the demodulation of the data symbols by solving
\begin{equation}\label{Detel}
\begin{split}
\hat{\mathbf{S}}=& \arg \min_{\mathbf{S}\in {\cal M}^{n_t \times K}} \|\mathbf{S}-(\check{\mathbf{S}} + \hat{\mathbf{E}})\|_F.
\end{split}
\end{equation}
\subsubsection{GSM System}
For the GSM system, we first demodulate the GSM signal $\mathbf{X}_t$ directly, i.e.,
\begin{equation}\label{Detell}
\begin{split}
\check{\mathbf{X}}_t =& \arg \min_{\mathbf{X}_t\in \{\mathcal{M}\cup\{0\}\}^{N_t \times K}} \|\mathbf{Y}_t - \mathbf{W}_t^H \hat{\mathbf{H}}_{t-1} \mathbf{X}_t \|_F,
\end{split}
\end{equation}
such that each column of $\check{\mathbf{X}}_t$ has $n_t$ non-zero elements. Denote
%\begin{equation}
$\mathbf{E}_t = \mathbf{X}_t - \check{\mathbf{X}}_t$.
%\end{equation}
Then (\ref{Re28}) becomes
\begin{equation}\label{DataGSM}
\mathbf{Y}_t = \mathbf{W}_t^H \mathbf{H}_t (\check{\mathbf{X}}_t + \mathbf{E}_t) + \mathbf{Q}_t.
\end{equation}

Similarly to (\ref{DSDPHB}), after dropping the subscript $t$, we have the following optimization problem for the GSM system,
which is also non-convex:
\begin{equation}\label{DSDPGSM}
\begin{aligned}
(\hat{\mathbf{h}},\hat{\mathbf{e}}) = & \arg \min \limits_{\tilde{\mathbf{h}}\in \mathbb{C}^{N_tN_r\times 1},\mathbf{e}\in \mathbb{C}^{KM_s\times 1}\hfill\atop \mathbf{V}\in \mathbb{C}^{(2N_t-1)\times(2N_r-1)},\varepsilon\in \mathbb{R}}
\frac{1}{2}\left\| {{{\tilde{\mathbf{y}}}} - ((\check{\mathbf{X}}^T+[\mathrm{vec}^{-1}(\mathbf{e})]^T)\otimes \mathbf{W}^H){{\tilde{\mathbf{h}}}}} \right\|_2^2  + \frac{\mu}{2N_tN_r}\mathrm{Tr}({\mathcal T}({\mathbf{V}})) \\& + \frac{\mu \varepsilon}{2} + \lambda \|\mathbf{e}\|_1\\
\rm{s.t.}~~~&\left[{\begin{array}{*{20}{c}}
{{\mathcal T}({\mathbf{V}})}&{\tilde{\mathbf{h}}}\\
{{\tilde{\mathbf{h}}}^H}&{\varepsilon}
\end{array}} \right] \succeq 0.
\end{aligned}
\end{equation}
Given the solution $\hat{\mathbf{h}}$ and $\hat{\mathbf{e}}$ to (\ref{DSDPGSM}), we obtain the estimates of the channel matrix $\hat{\mathbf{H}}={\mathrm{vec}}^{-1}(\hat{\mathbf{h}})$ and the sparse error matrix $\hat{\mathbf{E}}={\mathrm{vec}}^{-1}(\hat{\mathbf{e}})$. Finally the demodulation of the data symbols is refined by solving
\begin{equation}
%\begin{split}
\hat{\mathbf{X}} = \arg\min_{\mathbf{X}\in \{\mathcal{M}\cup\{0\}\}^{N_t \times K},
~\|\check{\mathbf{X}}(:,k)\|_0=n_t, k=1,2, ..., K} \|\mathbf{X}-(\check{\mathbf{X}} + \hat{\mathbf{E}})\|_F.
%\end{split}
\end{equation}

We name the channel estimators given by (\ref{DSDPHB}) and (\ref{DSDPGSM}) as the data-aided estimators based on atomic norm and $\ell_1$-norm for the HB system (Atom-DA-HB) and GSM system (Atom-DA-GSM), respectively.

\section{Efficient Non-convex Algorithms}

In the previous section, the proposed pilot-assisted channel estimator (\ref{PSDP}) is an SDP, which can be solved by off-the-shelf solvers such as SeDuMi \cite{SeDuMi} and SDPT3 \cite{SDPT3}. However, these solvers tend to be slow, especially for high-dimensional problems. Even though it is possible to develop a more efficient iterative algorithm based the alternating direction method of multipliers (ADMM) \cite{ADMM}, it needs to perform eigenvalue decomposition at each iteration, entailing a computational complexity $\mathcal{O}(N_t^3N_r^3)$, again posing a complexity issue for large-scale problems. Moreover, the data-aided channel estimators in (\ref{DSDPHB}) and (\ref{DSDPGSM}) are non-convex and hence efficient solvers need to be developed. In this section, we develop efficient non-convex solvers for (\ref{PSDP}), (\ref{DSDPHB}) and (\ref{DSDPGSM}). First, we take the Atom-DA-HB case as an example to derive the proposed non-convex solver, and then the non-convex solver is directly extended to the Atom-pilot and Atom-DA-GSM cases. Note that for the Atom-pilot case, the proposed non-convex solver has a much lower complexity than the convex counterpart, at the expense of slight performance degradation.

\subsection{Non-convex Factorization}

According to Lemma 2 in \cite{Le2}, suppose $\mathbf{\tilde{h}}=\sum\limits_{l=1}^L \alpha_l(\mathbf{a}_R(\phi_l)\otimes\mathbf{a}_T^H(\theta_l))$ is the solution to (\ref{DSDPHB}). If $\bar N = \min(N_t, N_r)\geq 1025$ and $\Delta = \inf\limits_{m \neq n} \sup\left\{|\theta_m - \theta_n|, |\phi_m - \phi_n | \right\} \geq \frac{4.76}{\bar N -1}$, then $\mathbf{\Psi}$ given by (\ref{DSDPHB}) satisfies rank$(\mathbf{\Psi}) = L$. It is worth noting that the condition $\bar N\geq 1025$ is a technical requirement that originally comes from Theorem 1.3 of \cite{Origin}. In Fig. 2, we illustrate via simulations that rank$(\mathbf{\Psi})=L$ holds even for small $\bar N$ as long as $(\bar N-1)\Delta$ is larger than a certain threshold.
\begin{figure}[!htb]
\begin{center}
\includegraphics[scale=0.6]{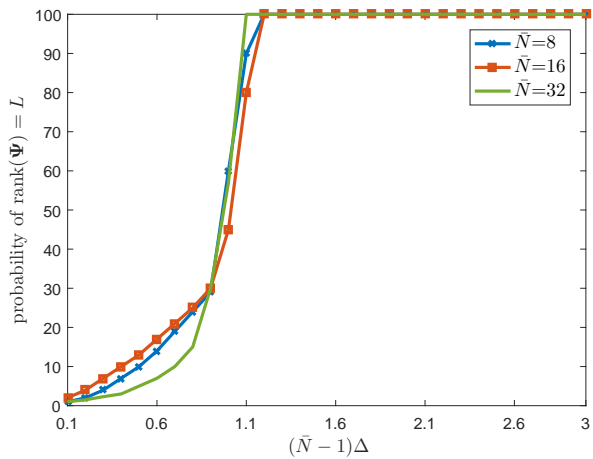}
%\captionsetup{font=small}
\caption{The probability that rank$(\mathbf{\Psi})=L$ vs. $(\bar N-1)\Delta$. We set $L=4$ in the simulations.}
\end{center}
\end{figure}

Hence, given the upper bound on the number of paths, say $\bar{L}$,
we can introduce the constraint of rank($\mathbf{\Psi}$) $\leq \bar{L}$ into (\ref{DSDPHB}),
whereby reducing the dimension of the positive semidefinite matrix in (\ref{DSDPHB}) to $(N_tN_r+1) \times \bar{L}\ll {(N_tN_r+1)}^2$.

In particular, we introduce the following non-convex factorization \cite{Non-convex}. Let $\mathbf \Psi = \mathbf \Gamma \mathbf \Gamma^H$, with $\mathbf \Gamma = [\mathbf \Gamma_0^T~~\mathbf \Gamma_1^T]^T \in \mathbb{C}^{(N_tN_r+1)\times \bar{L}}$, $\mathbf{\Gamma}_0 \in \mathbb{C}^{N_tN_r\times \bar{L}}$ and $\mathbf{\Gamma}_1 \in \mathbb{C}^{1\times \bar{L}}$, then we have ${\mathcal T}({\mathbf{V}}) = \mathbf \Psi_0 = \mathbf \Gamma_0 \mathbf \Gamma_0^H$, $\mathbf{\tilde h} = \mathbf{\Gamma}_0 \mathbf{\Gamma}_1^H$ and $\varepsilon = \mathbf{\Gamma}_1 \mathbf{\Gamma}_1^H$. This way the constraints ${\mathbf{\Psi}}\succeq 0$ in (\ref{DSDPHB}) and rank$(\mathbf{\Psi}) \leq \bar{L}$ are both satisfied.
Moreover since ${\mathcal T}({\mathbf{V}})=\mathbf{\Psi}_0$, the constraint $\mathbf{\mathcal{P}}_{\mathbf{\mathcal{T}}}(\mathbf{\Psi_0})=\mathbf{\Psi}_0$ need to be imposed, where $\mathbf{\mathcal{P}}_\mathbf{\mathcal{T}}(\cdot)$ denotes the projection of the input matrix onto a block Toeplitz matrix defined as the same as (\ref{Toeplitz}). Specifically, let $\mathbf{\mathcal{P}}_{\mathbf{\mathcal{T}}}(\mathbf{\Psi}_0)= \mathbf{\mathcal{T}}(\mathds{G}(\mathbf{\Psi}_0))$, where $G(\cdot)$ outputs an $(2N_t-1)\times(2N_r-1)$ matrix with an $N_tN_r\times N_tN_r$ matrix input $\mathbf{\Psi}_0$. In particular, if we partition $\mathbf{\Psi}_0$ into $N_r \times N_r$ blocks, i.e.,
\begin{eqnarray}
\mathbf{\Psi}_0 = \left[ {\begin{array}{*{20}{c}}
{{\mathbf{D}_{1,1}}}&{{\mathbf{D}_{1,2}}}& \cdots &{{\mathbf{D}_{1,{N_r}}}}\\
{{\mathbf{D}_{2,1}}}&{{\mathbf{D}_{2,2}}}& \cdots &{{\mathbf{D}_{2,{N_r}}}}\\
 \vdots & \vdots & \ddots & \vdots \\
{{\mathbf{D}_{{N_r},1}}}&{{\mathbf{D}_{{N_r},2}}}& \cdots &{{\mathbf{D}_{{N_r},{N_r}}}}
\end{array}} \right] \in \mathbb{C}^{N_tN_r\times N_tN_r},
\end{eqnarray}
with the $(p,q)$-th element of $\mathbf{D}_{m,n}$ denoted as $d_{p,q}^{m,n}$, $p,q=1,2,...,N_t; m,n=1,2,...,N_r$, then the $(i,j)$-th element of $\mathds{G}(\mathbf{\Psi}_0)$ is $\frac{\sum \limits_{p-q=i}^{m-n=j} d_{p,q}^{m,n}}{\kappa_{i,j}}$, $\kappa_{i,j} = (N_t-|i|)(N_r-|j|)$, $i=-N_t+1,-N_t+2,...,N_t-1,j=-N_r+1,-N_r+2,...,N_r-1$.
If we partition $\mathbf{\mathcal{T}}(\mathds{G}(\mathbf{\Psi}_0))$ into $N_r \times N_r$ blocks, e.g.,
$\mathbf{\tilde{D}}_{m,n} \in \mathbb{C}^{N_t\times N_t}$ with the $(p,q)$-th element of
$\mathbf{\tilde D}_{m,n}$ denoted as $\tilde d_{p,q}^{m,n}$, then
$\tilde d_{p,q}^{m,n} = \frac{\sum \limits_{p-q=i}^{m-n=j} d_{p,q}^{m,n}}{\kappa_{i,j}}$.
Therefore, the problem defined in (\ref{DSDPHB}) can be transformed into
\begin{equation}\label{Pro1}
\begin{aligned}
&\arg \min \limits_{\mathbf{\Gamma}\in \mathbb{C}^{N_tN_r\times \bar L}\hfill\atop \mathbf{e}\in \mathbb{C}^{KM_s\times 1}}
\frac{\mu}{2N_tN_r}\mathrm{Tr}(\mathbf{\Psi}_0) + \frac{1}{2}\left\| {{{\tilde{\mathbf{y}}}} - ((\check{\mathbf{S}}^T+[\mathrm{vec}^{-1}(\mathbf{e})]^T)\mathbf{F}^T \otimes \mathbf{W}^H) {{\tilde{\mathbf{h}}}}} \right\|_2^2 + \frac{\mu \varepsilon}{2} + \lambda \|\mathbf{e}\|_1\\
&\rm{s.t.}~~~~\mathbf{\mathcal{T}}(\mathds{G}(\mathbf{\Psi}_0))=
\mathbf{\Psi}_0.
\end{aligned}
\end{equation}

\subsection{Conjugate Gradient Descent Algorithm}

To solve (\ref{Pro1}), we first transform it into a smooth unconstrained optimization problem and then apply the CGD method to solve it. In particular we replace the constraint $\mathbf{\mathcal{P}}_{\mathbf{\mathcal{T}}}(\mathbf{\Psi}_0)=
\mathbf{\Psi}_0$ with the penalty term $\frac{\varrho}{2}\|\mathbf{\mathcal{P}}_{\mathbf{\mathcal{T}}}(\mathbf{\Psi}_0)
-\mathbf{\Psi}_0\|_F^2$ in the objective function. Moreover, since $\|\cdot\|_1$ in the objective function is non-smooth, we approximate it with
\begin{equation}\label{mu}
\|\mathbf{e}\|_1 \approx \psi_\tau(\mathbf{e})= \tau \sum\limits_{m=1}^{KM_s} \log \cosh(|e_m|/\tau),
\end{equation}
where $\mathbf{e}=[e_1,e_2,...,e_{KM_s}]^T$ and the parameter $\tau$ controls the smoothing level as illustrated in Fig. 3.
\begin{figure}[!htb]
\begin{center}
\includegraphics[scale=0.6]{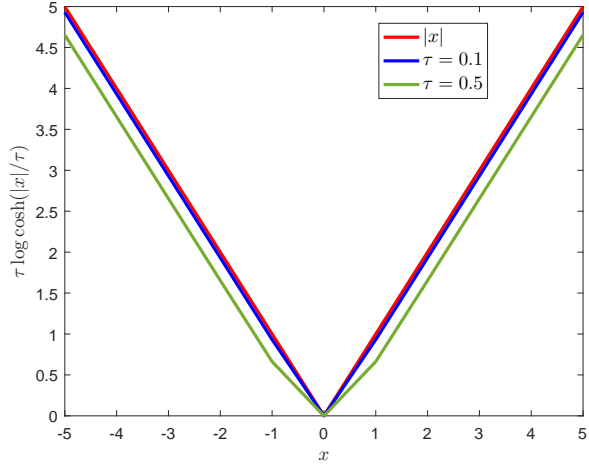}
%\captionsetup{font=small}
\caption{The smooth $\ell_1$-norm surrogate defined in (\ref{mu}). The surrogate approximates the $\ell_1$-norm more closely when $\tau$ gets smaller.}
\end{center}
\end{figure}
Hence, (\ref{Pro1}) is replaced by the following unconstrained optimization problem
\begin{equation}\label{Pro}
\min\limits_{\mathbf{\Gamma}\in \mathbb{C}^{(N_tN_r+1)\times \bar{L}}\hfill\atop\mathbf{e}\in \mathbb{C}^{KM_s \times 1}}~~ \zeta(\mathbf{\Gamma}\mathbf{\Gamma}^H,\mathbf{e}),
\end{equation}
where
\begin{equation}\label{Min}
\begin{aligned}
\zeta(\mathbf{\Gamma}\mathbf{\Gamma}^H,\mathbf{e})&= \frac{\mu}{2N_tN_r}\mathrm{Tr}(\mathbf{\Psi}_0) + \frac{\mu \varepsilon}{2} + \frac{1}{2}\left\| {{{\tilde{\mathbf{y}}}} - (\underbrace{(\check{\mathbf{S}}^T+[\mathrm{vec}^{-1}(\mathbf{e})]^T)\mathbf{F}^T \otimes \mathbf{W}^H}_{\mathbf \Theta}) {\tilde{\mathbf{h}}}} \right\|_2^2 \\
&+\frac{\varrho}{2}\|\mathbf{\mathcal{T}}(\mathds{G}(\mathbf{\Psi}_0))-\mathbf{\Psi}_0\|_F^2 + \lambda \psi_\tau(\mathbf{e}).
\end{aligned}
\end{equation}
The CGD algorithm \cite{CG} for solving (\ref{Min}) performs the following iterations
\begin{eqnarray}\label{Upd1}
\mathbf{\Gamma}^{\ell}&=&\mathbf{\Gamma}^{\ell-1}+\varsigma^{\ell}\mathbf{B}^{\ell},\\
\mathbf{e}^{\ell}&=&\mathbf{e}^{\ell-1}+\varsigma^{\ell}\mathbf{b}^{\ell},
\end{eqnarray}
where $\varsigma^{\ell}$ is the step size, $\mathbf{b}^{\ell}$ and $\mathbf{B}^{\ell}$ are the search directions at step $\ell$, evaluated as the weighted sum of the gradient at present iteration and the search direction used at the previous iteration. Specifically, let $\nabla_{\mathbf{\Gamma}}^{\ell}\zeta$ and $\nabla_{\mathbf{e}}^{\ell}\zeta$ be the gradients of $\zeta(\mathbf{\Gamma}\mathbf{\Gamma}^H,\mathbf{e})$ at the $\ell$-th iteration, then we have
\begin{eqnarray}\label{Upd2}
\mathbf{B}^{\ell} &=& -\nabla_{\mathbf{\Gamma}}^{\ell}\zeta+\omega^{\ell}\mathbf{B}^{\ell-1},\\
\mathbf{b}^{\ell} &=& -\nabla_{\mathbf{e}}^{\ell}\zeta+\omega^{\ell}\mathbf{b}^{\ell-1},
\end{eqnarray}
where
\begin{equation}\label{End}
\omega^{\ell} = \frac{\langle \nabla_{\mathbf{\Gamma}}^{\ell}\zeta, \mathbf{R}^{\ell} \rangle + \langle \nabla_{\mathbf{e}}^{\ell}\zeta, \mathbf{r}^{\ell} \rangle}{\langle \mathbf{B}^{\ell-1}, \mathbf{R}^{\ell} \rangle + \langle \mathbf{b}^{\ell-1}, \mathbf{r}^{\ell} \rangle},
\end{equation}
with $\langle \mathbf{A},\mathbf{C}\rangle$ being defined as $\langle \mathbf{A},\mathbf{C}\rangle = \mathrm{Tr}(\mathbf{C}^H \mathbf{A})$,
and
\begin{eqnarray}
\mathbf{R}^{\ell} &=& \nabla_{\mathbf{\Gamma}}^{\ell}\zeta-\nabla_{\mathbf{\Gamma}}^{\ell-1}\zeta,\\
\mathbf{r}^{\ell} &=& \nabla_{\mathbf{e}}^{\ell}\zeta-\nabla_{\mathbf{e}}^{\ell-1}\zeta.
\end{eqnarray}
%\begin{eqnarray}
%\mathbf{R}_{\ell-1}&=&\nabla_{\mathbf{\Gamma}}^{\ell}\zeta-\nabla_{\mathbf{\Gamma}}^{(\ell-1)}\zeta,\\ \mathbf{r}_{\ell-1}&=&\nabla_{\mathbf{e}}^{\ell}\zeta-\nabla_{\mathbf{e}}^{(\ell-1)}\zeta.
%\end{eqnarray}
The expressions of the gradients $\nabla_{\mathbf{\Gamma}}^{\ell}\zeta$ and $\nabla_{\mathbf{e}}^{\ell}\zeta$ are derived in Appendix.

Note that the above CGD algorithm can also be used to solve the data-aided channel estimation problem for the GSM system, by replacing $\mathbf \Theta$ in (\ref{Min}) with $(\check{\mathbf{X}}^T+\mathbf{E}^T)\otimes\mathbf{W}^H$.
Moreover, the CGD algorithm can be used to solve the pilot-assisted channel estimation problem in (27) as well, i.e., using $\mathbf{e}=\mathbf{0}_{KM_s \times 1}$, $\mathbf{\Theta}=(\mathbf{S}^T\mathbf{F}^T)\otimes\mathbf{W}^H$ for the HB system and $\mathbf{\Theta}=\mathbf{X}^T\otimes\mathbf{W}^H$ for the GSM system,
respectively.

For clarity, we summarize the proposed non-convex solver for the Atom-DA-HB estimator in \textbf{Algorithm 1}, \textbf{Algorithm 2} and \textbf{Algorithm 3}, and the algorithms of the Atom-DA-GSM estimator and the Atom-pilot estimator are similar. To guarantee that the objective function
does not increase with $\ell$, the Armijo line search \cite{Armi} is employed (line 10 of \textbf{Algorithm 2} and line 11 of \textbf{Algorithm 3}), so that the algorithm converges to a stationary point of the surrogate problem, namely, the point where the smoothed objective function (\ref{Min}) has vanishing gradient.

\begin{algorithm}[!htb] %算法的开始
\renewcommand{\algorithmicrequire}{\textbf{Input:}}
\renewcommand\algorithmicensure {\textbf{Output:} }
\caption{Atom-DA-HB estimator} %算法的标题
\label{alg:Framwork} %给算法一个标签，这样方便在文中对算法的引用
\begin{algorithmic}[1] %这个1 表示每一行都显示数字
\REQUIRE $T$, $K$, $M_s$, $\mathbf{S}_0$ and $\{\mathbf{Y}_t\}_{t=0}^T$\\
\ENSURE $\hat{\mathbf{H}}=[\hat{\mathbf{H}}_1,\hat{\mathbf{H}}_2,...,\hat{\mathbf{H}}_T]$\\
\STATE $t=0$.
\STATE $\hat{\mathbf{h}}_{0}=\mathrm{CGPilot}(K,M_s,\mathbf{S}_0,\mathbf{F}_0,\mathbf{W}_0,\mathbf{Y}_0)$, $\hat{\mathbf{H}}_0 = \mathrm{vec}^{-1}(\hat{\mathbf{h}}_{0})$.
\STATE Set $\mathbf{F}_1$ and $\mathbf{W}_1$ using $\mathbf{\hat{H}}_0$ according to, e.g., \cite{Beamforming}.
\FOR{$t=1$ to $T$}
\STATE Obtain $\check{\mathbf{S}}_t$ using (\ref{Dete}).
\STATE $(\hat{\mathbf{h}}_{t},\hat{\mathbf{e}}_{t})=\mathrm{CGData}(K,M_s,\check{\mathbf{S}}_t,\mathbf{F}_t,\mathbf{W}_t,\mathbf{Y}_t)$.
%\STATE 2. Identify the non-zero entries $\hat{u}_{t,k}$ in $\hat{\mathbf{x}}_{t,k}$,
%then map $(\widehat{u}_{t,k},\widehat{\mathbf{x}}_{t,k}(\widehat{u}_{t,k}))$ back to the information bits
%$\widehat{\mathbf{s}}_{t,k}$ for GSM MIMO, or identify $\widehat{\mathbf{s}}_{t,k}$ according to $\widehat{\mathbf{x}}_{t,k} = \mathbf{F}_t \widehat{\mathbf{s}}_{t,k}$ for Hybrid MIMO.
\STATE $\hat{\mathbf{H}}_t = \mathrm{vec}^{-1}(\hat{\mathbf{h}}_{t})$, $\hat{\mathbf{E}}_t = \mathrm{vec}^{-1}(\hat{\mathbf{e}}_{t})$, and update $\hat{\mathbf{S}}_t$ using (\ref{Detel});
\STATE Update $\mathbf{F}_{t+1}$ and $\mathbf{W}_{t+1}$ using $\mathbf{\hat{H}}_t$ according to, e.g., \cite{Beamforming}.
%\STATE Obtain $\mathbf{Y}_t$ according to (7);
\ENDFOR
\end{algorithmic}
\end{algorithm}
\vspace{-4mm}
\begin{algorithm}[!htb] %算法的开始
\renewcommand{\algorithmicrequire}{\textbf{Input:}}
\renewcommand\algorithmicensure {\textbf{Output:} }
\caption{~~$\hat{\mathbf{h}}$ = CGPilot$(K,M_s,\mathbf{S}_0,\mathbf{F}_0,\mathbf{W}_0,\mathbf{Y}_0)$} %算法的标题
\label{alg:Framwork} %给算法一个标签，这样方便在文中对算法的引用
\begin{algorithmic}[1] %这个1 表示每一行都显示数字
\REQUIRE $\epsilon$, $K$, $M_s$, $\varrho$, $\mu$, $\mathbf{Y}_0$ and $\mathbf{S}_0$\\
\ENSURE $\hat{\mathbf{h}}$\\
\STATE $\ell=0$.\\
\STATE \textbf{Do}\\
\STATE $\ell\leftarrow \ell+1$.\\
\STATE Calculate $\nabla_{\mathbf{\Gamma}}^{\ell}\zeta$ using (\ref{Chain}), (\ref{A}), (\ref{Q}) and  (\ref{Deri}).\\
\IF {$\ell=1$}
\STATE $\mathbf{B}^{\ell}=-\nabla_{\mathbf{\Gamma}}^{\ell}\zeta$,\\
\ELSE
\STATE Calculate $\mathbf{B}^{\ell}$ using (\ref{Upd2}), (\ref{End})-(48).\\
\ENDIF\\
\STATE Update $\mathbf{\Gamma}^{\ell}$ using (\ref{Upd1}) with $\varsigma^{\ell}$ obtained via Armijo line search.\\
\STATE \textbf{While} {$\|\nabla_{\mathbf{\Gamma}}^{\ell}\zeta\|_F > \epsilon$}.\\
\STATE $\hat{\mathbf{h}}= \mathbf{\Gamma}_0^{\ell} (\mathbf{\Gamma}_1^{\ell})^H$ with $\mathbf{\Gamma}^{\ell} = [(\mathbf \Gamma_0^{\ell})^T~(\mathbf \Gamma_1^{\ell})^T]^T$.
\end{algorithmic}
\end{algorithm}
\vspace{-4mm}
\begin{algorithm}[!htb] %算法的开始
\renewcommand{\algorithmicrequire}{\textbf{Input:}}
\renewcommand\algorithmicensure {\textbf{Output:} }
\caption{~~$(\hat{\mathbf{h}},\hat{\mathbf{e}})$ = CGData$(K,M_s,\check{\mathbf{S}}_t,\mathbf{F}_t,\mathbf{W}_t,\mathbf{Y}_t)$} %算法的标题
\label{alg:Framwork} %给算法一个标签，这样方便在文中对算法的引用
\begin{algorithmic}[1] %这个1 表示每一行都显示数字
\REQUIRE $\epsilon$, $K$, $M_s$, $\varrho$, $\mu$, $\tau$, $\lambda$, $\mathbf{Y}_t$ and $\check{\mathbf{S}}_t$\\
\ENSURE $\hat{\mathbf{h}}$, $\hat{\mathbf{e}}$\\
\STATE $\ell=0$.\\
\STATE \textbf{Do}\\
\STATE $\ell=\ell+1$.\\
\STATE Calculate $\nabla_{\mathbf{e}}^{\ell}\zeta$ using (\ref{Derive}) and (\ref{E}).\\
\STATE Calculate $\nabla_{\mathbf{\Gamma}}^{\ell}\zeta$ using (\ref{Chain}), (\ref{A}), (\ref{Q}) and (\ref{Deri}).\\
\IF {$\ell=1$}
\STATE $\mathbf{B}^{\ell}=-\nabla_{\mathbf{\Gamma}}^{\ell}\zeta$, $\mathbf{b}^{\ell}=-\nabla_{\mathbf{e}}^{\ell}\zeta$,\\
\ELSE
\STATE Calculate $\mathbf{B}^{\ell}$ and $\mathbf{b}^{\ell}$ using (\ref{Upd2})-(48).\\
\ENDIF\\
\STATE Update $\mathbf{\Gamma}^{\ell}$ and $\mathbf{e}^{\ell}$ using (\ref{Upd1}) and (43) with $\varsigma^{\ell}$ obtained via Armijo line search.\\
\STATE \textbf{While} {$\|\nabla_{\mathbf{\Gamma}}^{\ell}\zeta\|_F > \epsilon$}.\\
\STATE $\hat{\mathbf{h}}= \mathbf{\Gamma}_0^{\ell} (\mathbf{\Gamma}_1^{\ell})^H$ with $\mathbf{\Gamma}^{\ell} = [(\mathbf \Gamma_0^{\ell})^T~(\mathbf \Gamma_1^{\ell})^T]^T$, $\hat{\mathbf{e}}=\mathbf{e}^{\ell}$.
\end{algorithmic}
\end{algorithm}

\section{Simulation Results}

In this section, we use simulations to illustrate the performance of the proposed algorithms in both the HB and GSM systems. The uniform linear arrays at the transmitter and receiver are equipped with $N_t=N_r=16$ antennas and $n_t=n_r=2$ RF chains, respectively. The channel matrix is generated according to (\ref{Chan}) where $\{\theta_l,\phi_l\}_{l=1}^L$ are uniformly generated within the interval of $[0,1)$ and the path amplitudes $\{\alpha_l\}_{l=1}^L$ are randomly generated according to distribution $\mathcal{CN}(0,\sigma_l^2)$ with equal variances, i.e., $\sigma_l^2=1, l = 1,2,...,L$. Following the setting of \cite{Channel-model2}, the average number of resolvable paths ranges from 1 to 8. The number of time blocks is $T=100$ and QPSK modulation is employed. The signal-to-noise ratio (SNR) is defined as $\frac{P}{N_t\sigma^2}$ with $P$ denoting the average transmission power. We use the normalized mean-square error (NMSE) defined as $10\log_{10}(\mathbb{E}[\|\mathbf{H}-\hat{\mathbf{H}}\|_F^2/\|\mathbf{H}\|_F^2])$ to evaluate the performance of channel estimation, and the symbol error rates (SERs) defined as $\mathbb{E}[\sum_{k=1}^K \eta(\mathbf{s}_k-\hat{\mathbf{s}}_k)/K]$ for the HB system and $\mathbb{E}[\sum_{k=1}^K \eta(\mathbf{x}_k-\hat{\mathbf{x}}_k)/K]$ for the GSM system with
$\eta(\mathbf{f},\hat{\mathbf{f}}) = 0$ if $\mathbf{f}=\hat{\mathbf{f}}$; else $\eta(\mathbf{f},\hat{\mathbf{f}}) = 1$,
to evaluate the performance of symbol demodulation.
%${\eta(\mathbf{f},\hat{\mathbf{f}}) = \left\{ \begin{gathered}
%0,~\mathbf{f}=\hat{\mathbf{f}}, \hfill \\
%1,~\mathbf{f}\neq \hat{\mathbf{f}}, \hfill
%\end{gathered} $

\subsection{Convergence of the Conjugate Gradient Descent Algorithm}
The computational complexity of the proposed CGD algorithm at each iteration is mainly
determined by the calculation of $\mathbf{\Gamma}\mathbf{\Gamma}^H$, whose complexity is
$\mathcal{O}(N_t^2N_r^2\bar{L})$. As $\bar{L}\ll N_t N_r$, the complexity per iteration is
much smaller than that of a classical eigenvalue decomposition, whereby, for large-dimensional
problems, the proposed non-convex approach can be faster than those based on the
first-order methods such as ADMM.
%For the cvx solver, primal-dual interior-point methods are usually used to solve SDP with the general complexity \cite{}.
We illustrate this fact through simulation examples, whose results are reported in Fig. 4. The parameters are SNR = $10$dB, $L = 5$ and $K = 8$. The non-convex solver is implemented by solving (\ref{Pro}) with the CGD algorithm. We compare the NMSE of the proposed algorithm with that given by solving (\ref{PSDP}) with the CVX \cite{CVX} and ADMM \cite{ADMM} solvers. As can be seen from Fig. 4, the results of the proposed algorithm is close to the solution given by the CVX after 300 iterations. Because the proposed algorithm runs much faster than the ADMM, it appears much more suitable for real-time implementation.

%The parameters are set as $L=2$, $\varrho = 5$ and the variance of the noise is $\sigma^2=0.01$. The non-convex solvers are implemented by solving (\ref{Pro}) with the CG algorithm. We compare the NMSEs of the channel estimators using the proposed algorithm with that given by solving (\ref{PSDP}), (\ref{Con1}) and (\ref{Con2}) using the CVX solver \cite{CVX}. Note that, for better performance evaluation of the proposed non-convex solver, the number of non-zero entries in $\mathbf{E}$ are controlled by the BER which is preset as 0.1.
\begin{figure}[!htb]
\centering
%\captionsetup{font=small}
  \includegraphics[width=8.5cm]{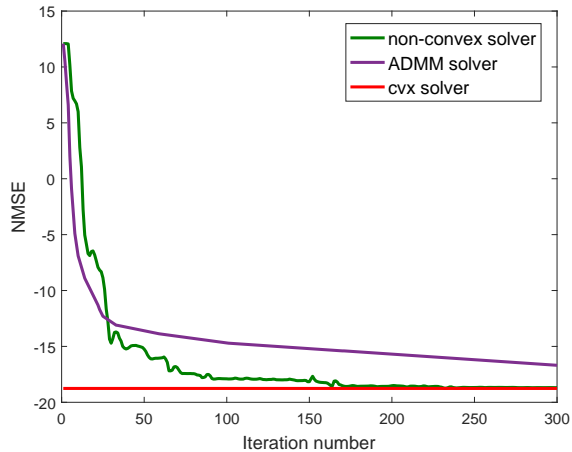}
\caption{Convergence behavior of the proposed non-convex solver. The non-convex solver takes 34 seconds with 300 iterations, the ADMM solver takes 142 seconds with 300 iterations, and the CVX solver of the Atom-pilot estimator takes 359 seconds in the HB system.}
\label{fig:res}
\end{figure}
%To simplify the expression, the results of Atom-pilot solved by SeDuMi and SDPT3 are omitted since they are visually identical to those by non-convex solver in the remainder of this paper.

\subsection{Pilot-assisted Channel Estimation}

To compared with the proposed Atom-pilot estimators, we consider two grid-based compressed sensing methods for performance comparison with the Atom-pilot estimators, i.e., the OMP and CS-L1 algorithms discussed in Section III, where the continuous parameter space $[0,1) \times [0,1)$ is discretized into a finite set of grids with $N_g$ grid points. For the Atom-pilot estimator, we use the CVX solver to solve (\ref{PSDP}), and the proposed non-convex solver to solve (\ref{Pro}) with $\mathbf{e}=\mathbf{0}_{KM_s\times 1}$, randomly initialized $\mathbf{\Gamma}^{0}$, $\bar{L}= 9$ and $\varrho=5$. The algorithm stops as the gradient norm is smaller than $\epsilon=0.01$. The weighting parameter is set as $\mu=\sigma \sqrt{N_tN_r\log(N_tN_r)}$ \cite{Atomic}.

During the pilot training stage, for the HB system, the analog filters $\mathbf{F}_A$ and $\mathbf{W}_A$ are designed according to (\ref{DFT}) and (9) respectively; for the GSM system, $\mathbf{W}_A$ is designed according to (9).
\begin{figure}[!htb]
%\begin{tabular}{cc}
%\captionsetup{font=small}
\begin{minipage}{0.5\linewidth}
  \centerline{\includegraphics[width=8.5cm]{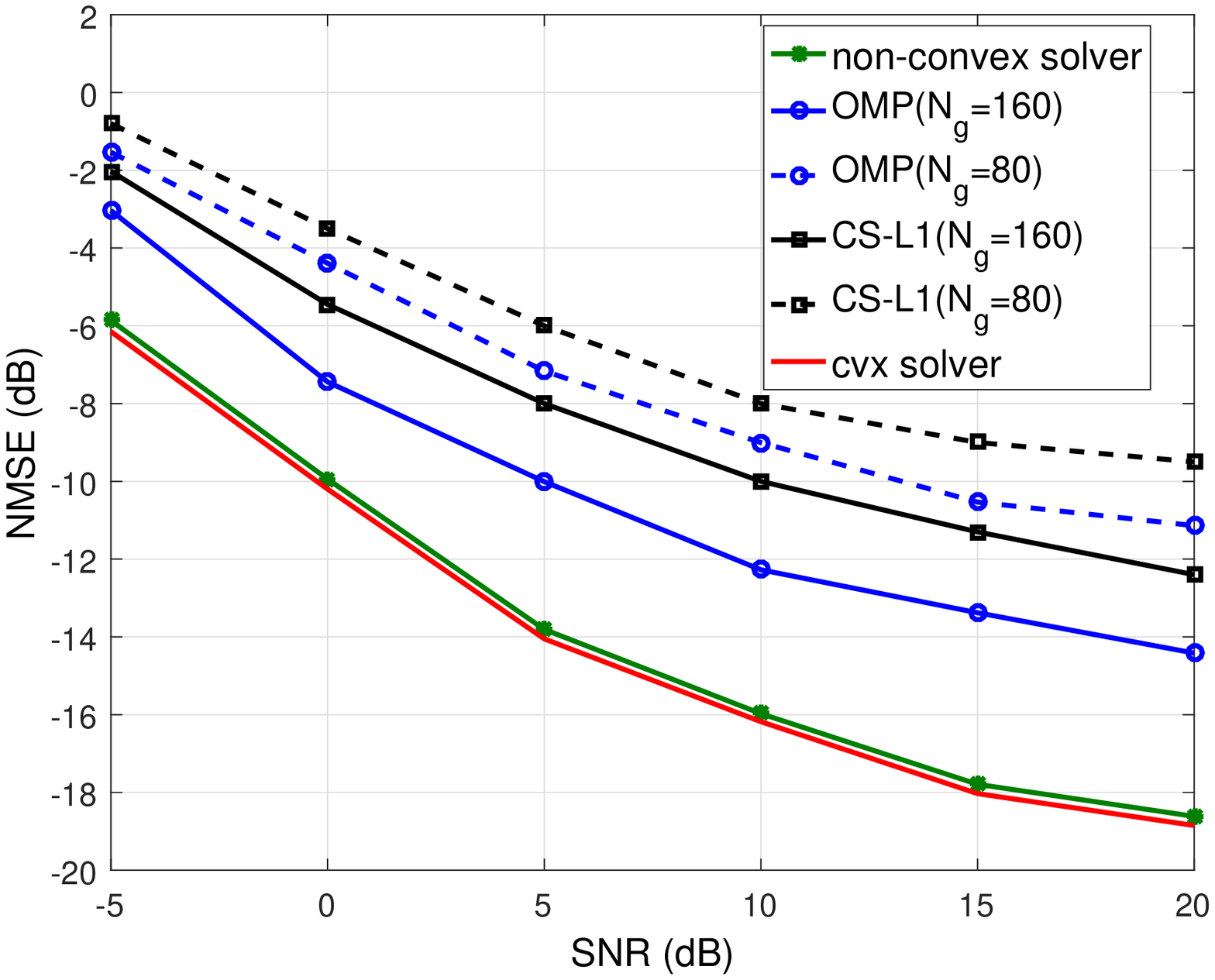}}
  \centerline{(a)}
\end{minipage}\hfill
\begin{minipage}{0.5\linewidth}
  \centerline{\includegraphics[width=8.5cm]{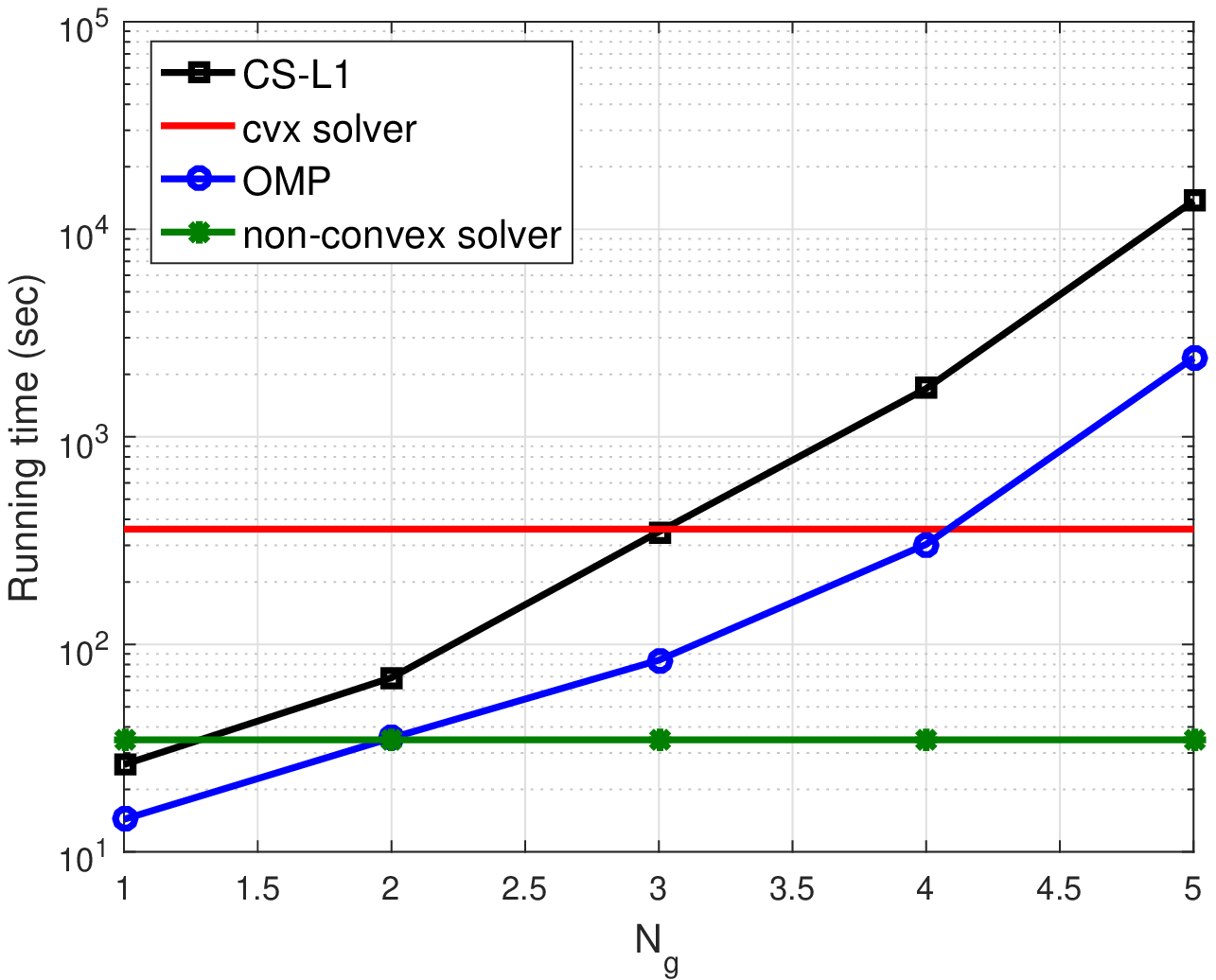}}
  \centerline{(b)}
\end{minipage}
\vfill
\begin{minipage}{0.5\linewidth}
  \centerline{\includegraphics[width=8.5cm]{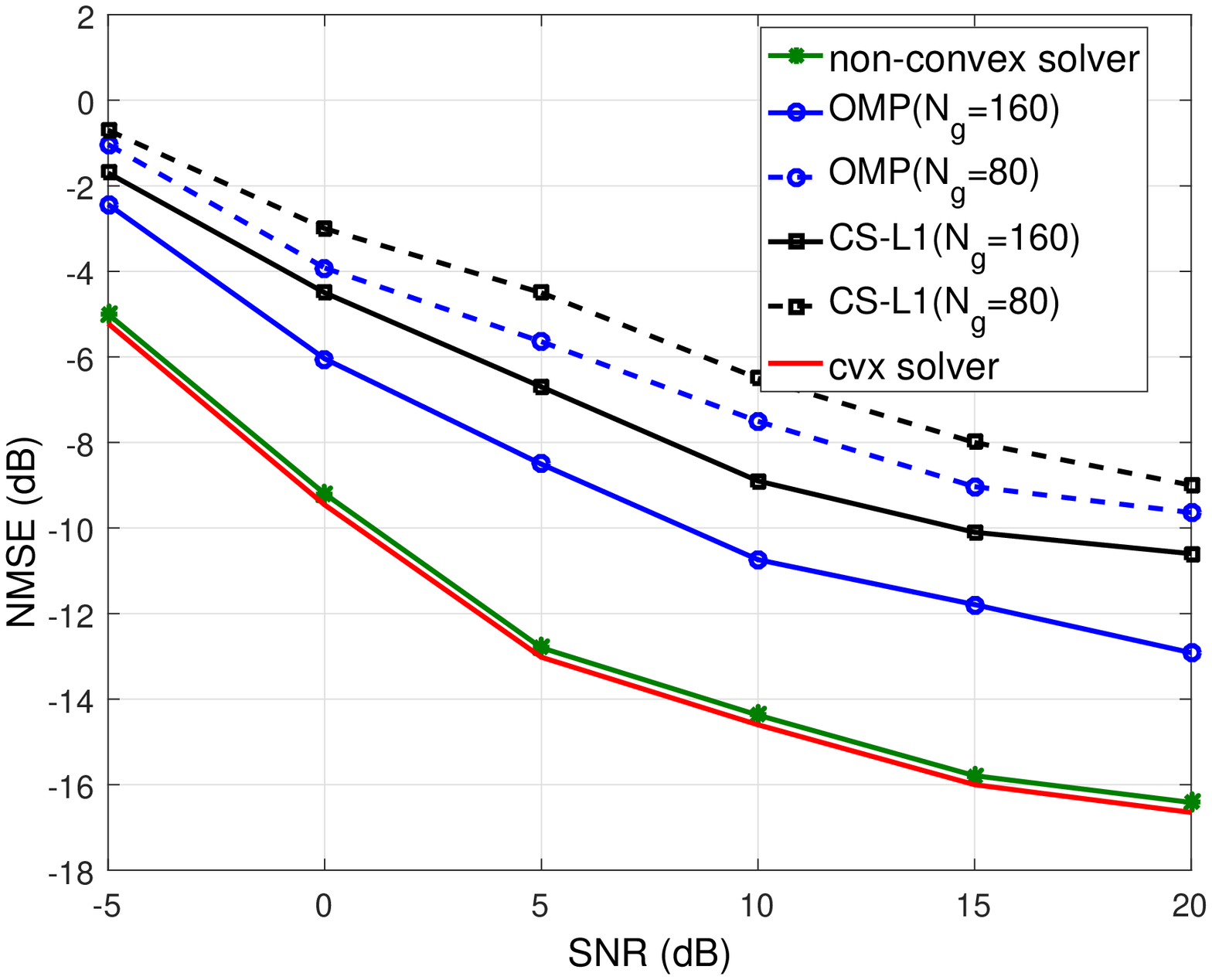}}
  \centerline{(c)}
\end{minipage}\hfill
\begin{minipage}{0.5\linewidth}
  \centerline{\includegraphics[width=8.5cm]{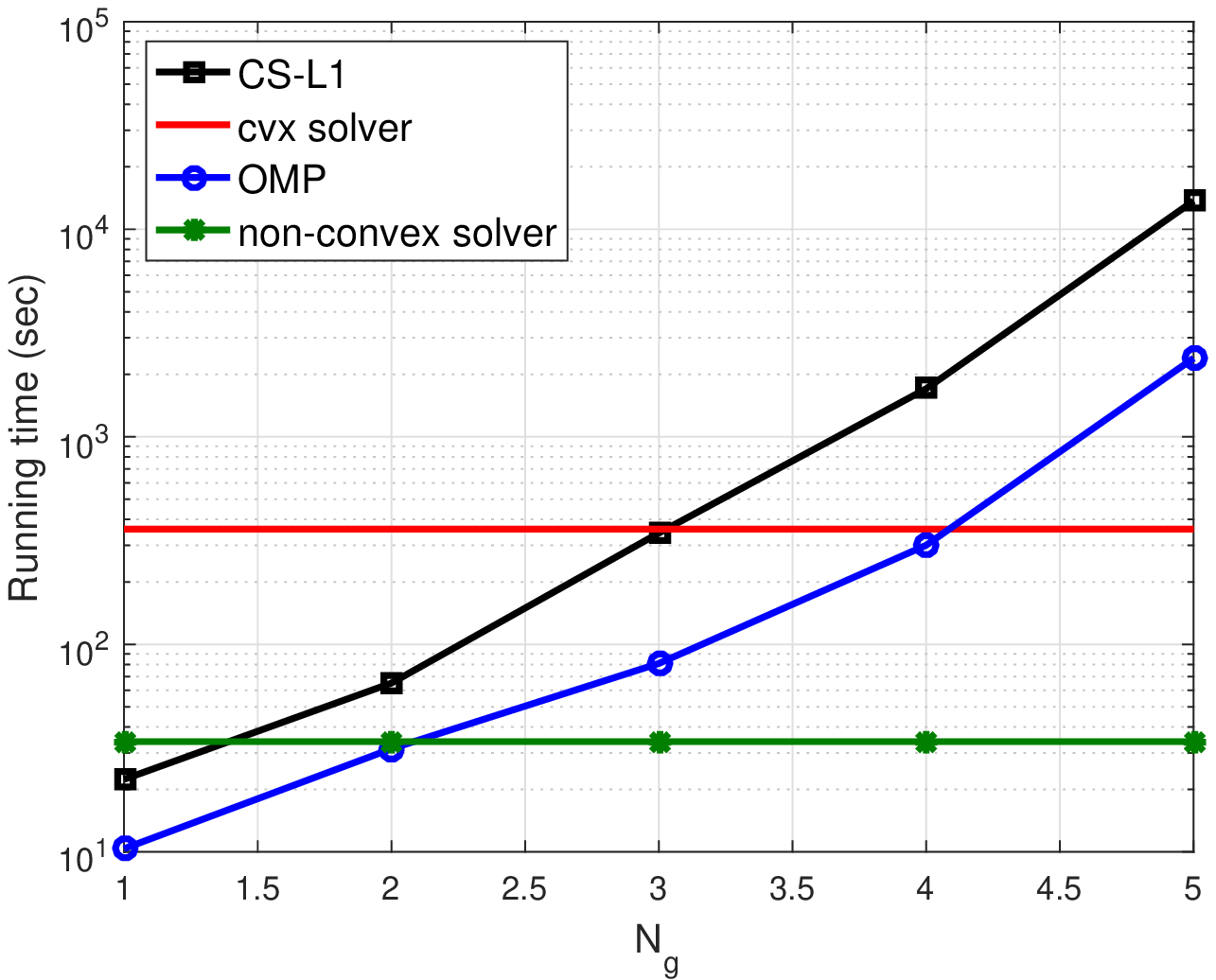}}
  \centerline{(d)}
\end{minipage}
%\end{tabular}
\caption{(a) NMSE performance of the Atom-pilot estimator in the HB system, $L$=3, $K$=8; (b) Running times for different estimators in the HB system, SNR=10dB; (c) NMSE performance of the Atom-pilot estimator in the GSM system, $L$=3, $K$=8; (d) Running times for different estimator in the GSM system, SNR=10dB.}
\label{fig:res}
\end{figure}

As shown in Fig. 5(a), under the same $N_g$, the accuracy of the OMP is always better than that of the CS-L1 algorithm, partly because the number of paths is very small, i.e., $L=3$, for which $\ell_0$-minimization usually results in better accuracy \cite{Le}. The OMP estimator is outperformed by the proposed Atom-pilot estimator, especially in the high SNR region. This is due to the fact that, both the CS-L1
and OMP estimators tend to mis-estimate the channel parameters due to basis mismatch.
When the grids become denser, the NMSE of the OMP and CS-L1 estimators become smaller. However, the computational complexity becomes higher as shown in Fig. 5(b). Fig. 5(c) and Fig. 5(d) show the performance comparisons in the GSM system. The results are similar to those of the HB system.

\subsection{Data-aided Channel Estimation}

Now we consider the performance of the proposed data-aided estimators with slowly time-varying channels. The channel matrices $\mathbf{H}_t, t=0,1,...,T$ are generated according to (\ref{Chan}) at the $t$-th time block. More specifically, for $t=0$, $\alpha_l^0$ are generated following complex Gaussian distribution, i.e., $\alpha_l^0 \sim \mathcal{CN}(0,\sigma_l^2)$, $\phi_l^0$ and $\theta_l^0$ are generated following uniform distribution, i.e., $\phi_l^0, \theta_l^0 \sim \mathrm{U}(-1,1)$ with $\mathrm{U}(a,b)$ denoting the uniform distribution in the interval $(a,b)$. To model the time correlation of the channel, at subsequent time blocks $(t=1,2,...,T)$, the variation of $\alpha_l^t$, $\phi_l^t$ and $\theta_l^t$ relative to $\alpha_l^{t-1}$, $\phi_l^{t-1}$ and $\theta_l^{t-1}$, i.e., $\Delta \alpha_l^t=\alpha_l^t-\alpha_l^{t-1}$, $\Delta \phi_l^t=\phi_l^t-\phi_l^{t-1}$ and $\Delta \theta_l^t=\theta_l^t-\theta_l^{t-1}$, follow distributions $\mathcal{CN}(0,0.01\sigma_l^2)$, $\mathrm{U}(-0.1+\phi_l^{t-1},0.1+\phi_l^{t-1})$ and $\mathrm{U}(-0.1+\theta_l^{t-1},0.1+\theta_l^{t-1})$, respectively. The weighting parameters for regularizing the sparse demodulation error are set as $\lambda=\mu/\sqrt{N_tN_r}$ and $\tau=0.01$. In this case for the Atom-pilot estimator, it estimates the channel based on the pilot at $t=0$ to obtain $\hat{\mathbf{H}}_0$ and uses it to demodulate the data for subsequent blocks $t=1,2,...,T$. The Atom-DA-HB and Atom-DA-GSM estimators, however, updates the channel matrix at each time block $t$ in a data-aided manner. We also simulate the case when the channel matrix in each block is estimated with pilot symbols which serves as the lower bound of the NMSE in channel estimation.

%To see the impact of unknown transmitted symbols, the channel estimation performance of Atom-DA-HB and the Atom-DA-GSM is also compared with the lower bound obtained through estimating the channel matrix with known pilot symbols in each time block.

%In the next, we will illustrate the performance of the Atom-DA-HB and the Atom-DA-GSM estimators and compare them with the Atom-pilot which estimates the channel matrix with known pilot symbols during $T$ time blocks, and the Atom-Data estimator which estimates the channel matrix with estimated symbols using atomic norm minimization, and demodulates the data symbols with estimated channel matrix using ML method during $T-1$ time blocks.
\begin{figure}[!htb]
%\begin{tabular}{cc}
\begin{minipage}{0.5\linewidth}
  \centerline{\includegraphics[width=8.5cm]{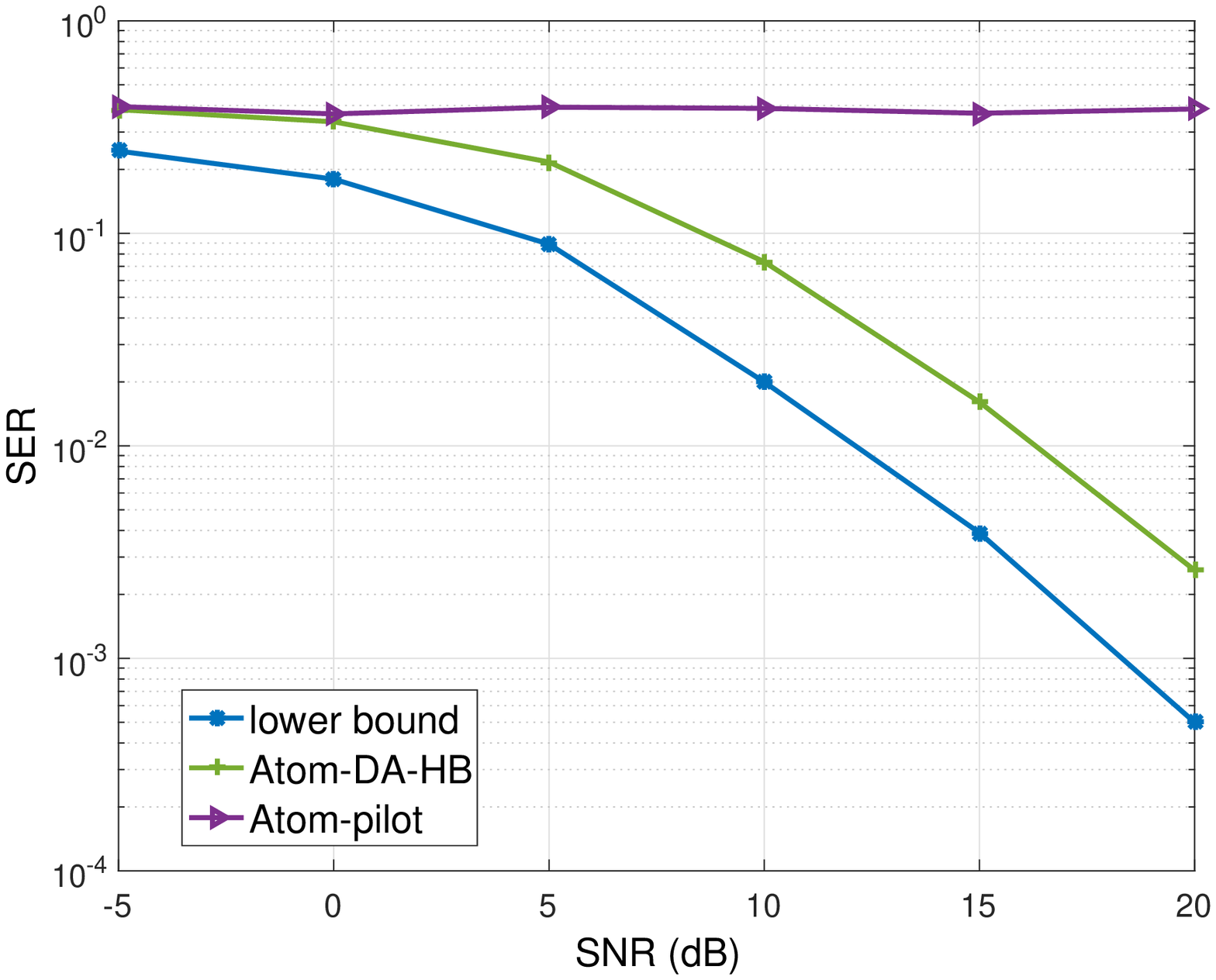}}
  \centerline{(a)}
\end{minipage}\hfill
\begin{minipage}{0.5\linewidth}
  \centerline{\includegraphics[width=8.5cm]{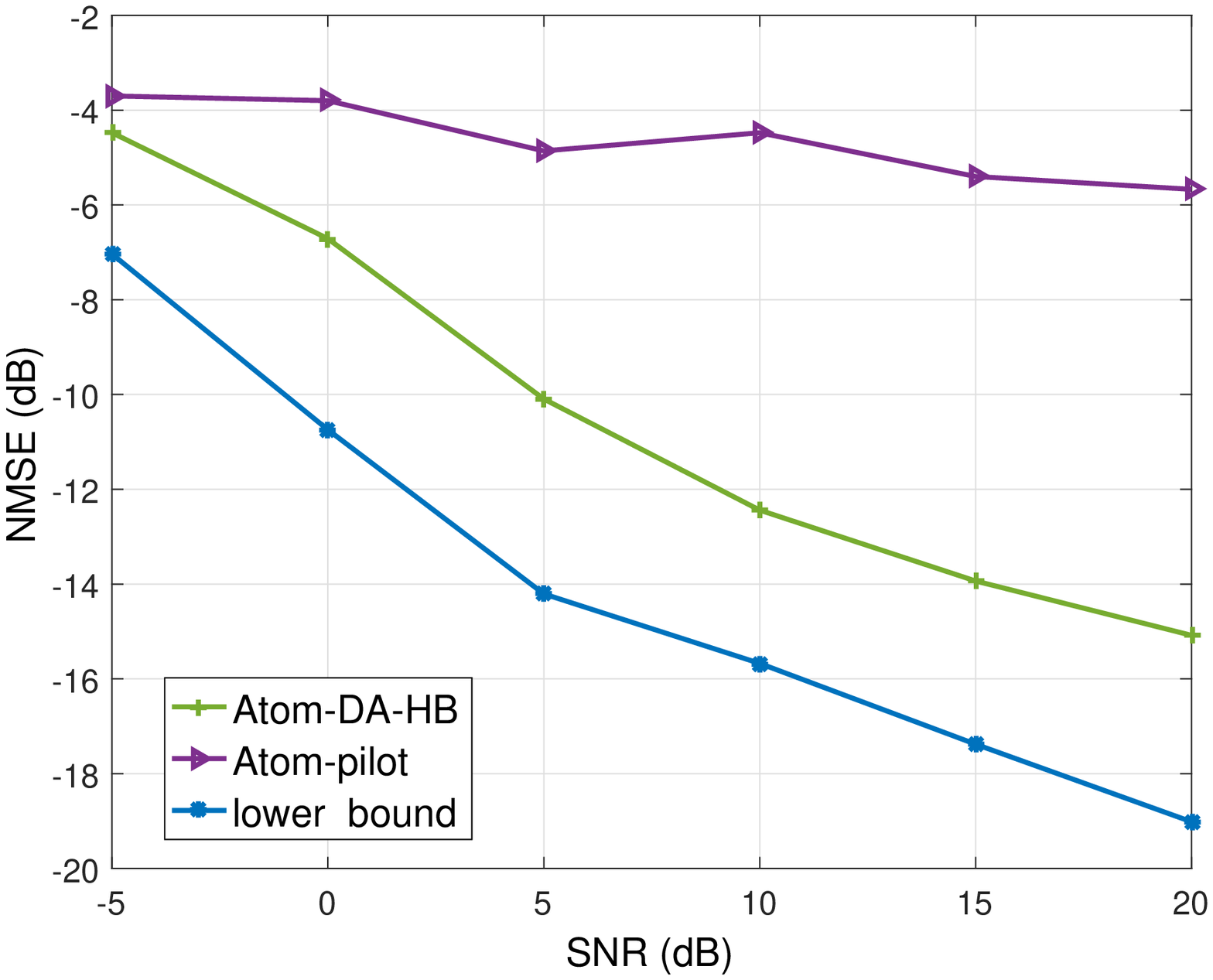}}
  \centerline{(b)}
\end{minipage}
\vfill
\begin{minipage}{0.5\linewidth}
  \centerline{\includegraphics[width=8.5cm]{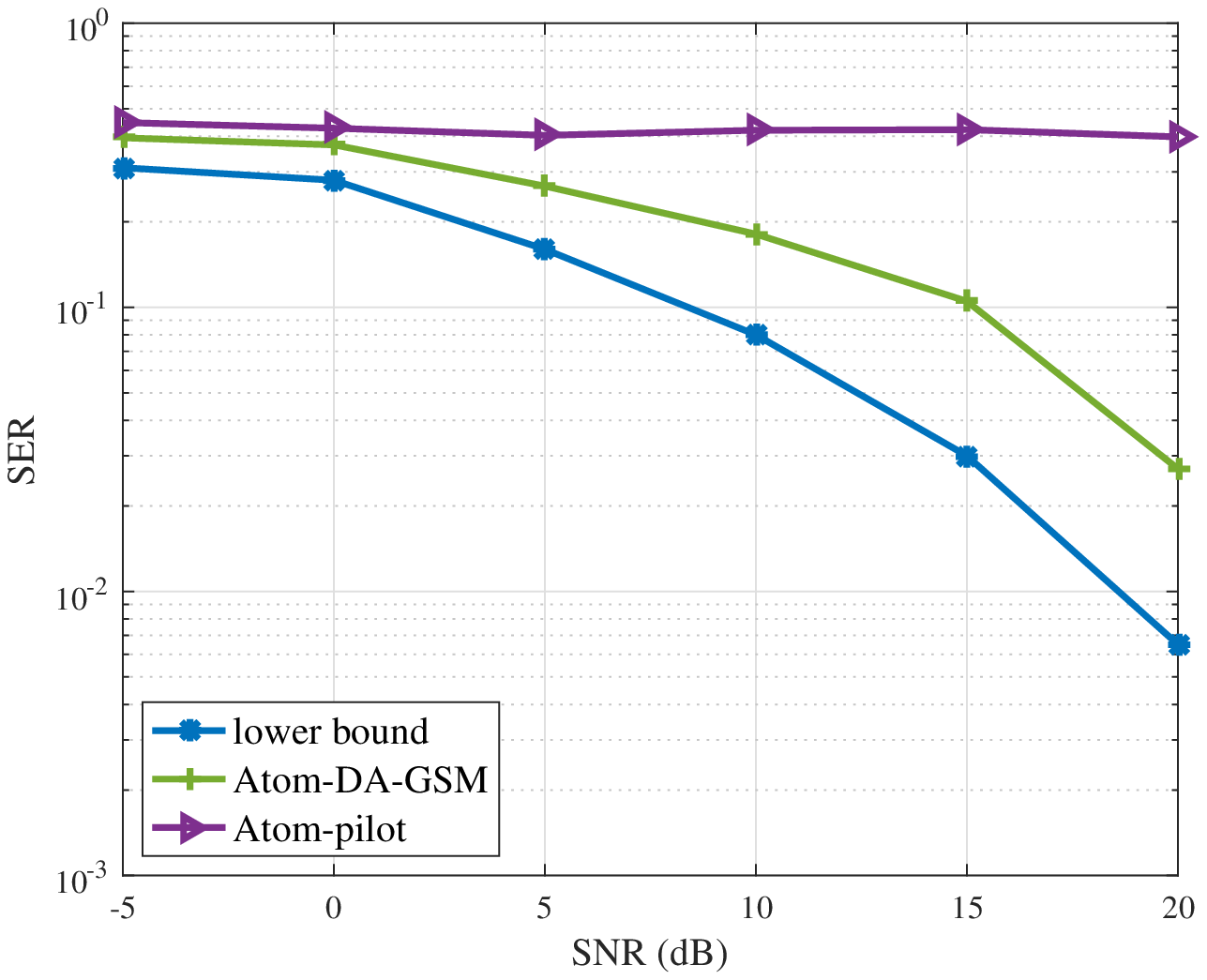}}
  \centerline{(c)}
\end{minipage}\hfill
\begin{minipage}{0.5\linewidth}
  \centerline{\includegraphics[width=8.5cm]{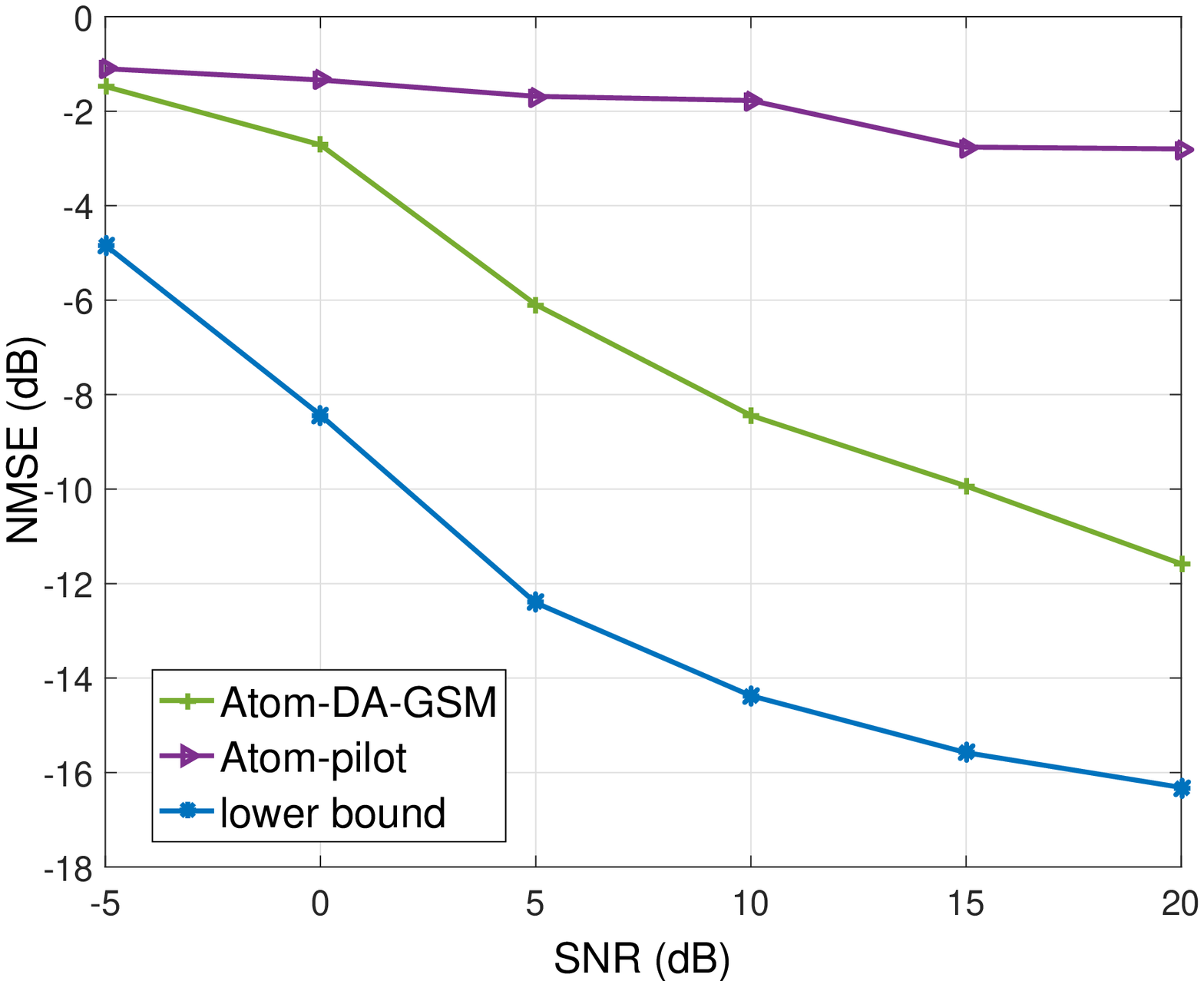}}
  \centerline{(d)}
\end{minipage}
%\end{tabular}
%\captionsetup{font=small}
\caption{Comparison of the estimators performance in the HB system: (a) SER, $L$=8, $K$=8; (b) NMSE, $L$=8, $K$=8; Comparison of the estimators performance in the GSM system: (c) SER, $L$=8, $K$=8; (d) NMSE, $L$=8, $K$=8.}
\label{fig:res}
\end{figure}

According to Fig. 6(a) and Fig. 6(b), the Atom-pilot estimator cannot work even at high SNR regions. Though there is still a gap of performance between the Atom-DA-HB estimator and the lower
bound, the Atom-DA-HB estimator always significantly outperforms the Atom-pilot estimator because it keeps tracking of the time-varying channel dynamics by estimating the channel and demodulation error alternately in each time block. For the Atom-DA-HB estimator, the NMSE curve decreases monotonically with the increasing SNR. This is because that the Atom-DA-HB estimator has better performance on channel estimation with high SNRs; on the other hand, the ML decoder are getting better when the SNR increases, and the demodulation error is getting
sparser, in which case the $\ell_1$-norm regularization of the Atom-DA-HB estimator has better performance on demodulation error estimation.
We plot both the SER and NMSE performance comparisons of the GSM system in Fig. 6(c) and Fig. 6(d). The results are similar to that of the HB system and the proposed Atom-DA-GSM estimator always performs better than the Atom-pilot estimator.

\section{Conclusions}
In this paper, we have proposed super-resolution channel estimators for HB-based and GSM-based mmWave systems. For the pilot-assisted scenarios, the proposed channel estimators are based on the atomic norm minimization that exploits the channel sparsity in the continuous angles of departure and arrival. For the data-aided scenario, the proposed channel estimator are based on both atomic norm minimization and $\ell_1$-minimization to exploit the sparsity in both channel and demodulation error. We have developed computationally efficient non-convex methods based on CGD to solve the formulated channel estimators. Simulation results indicate that the proposed algorithms outperform the on-grid CS channel estimators. Moreover, the proposed data-aided estimators can effectively track the time-varying channel dynamics.
%\appendices
\appendix
\subsection{Gradient Calculations}
%\vspace{-1.6em}
%For notational simplicity we drop the superscript $\ell$ during the derivations of $\nabla_{\mathbf{\Gamma}}^{\ell}\zeta$ and $\nabla_{\mathbf{e}}^{\ell}\zeta$.

Firstly, to derive $\nabla_{\mathbf{\Gamma}}\zeta$, we rewrite (\ref{Min}) as
\begin{equation}
%\begin{aligned}
\zeta(\mathbf{\Gamma}\mathbf{\Gamma}^H,\mathbf{e})= \frac{\mu}{2N_tN_r}\mathrm{Tr}(\mathbf{\Psi}_0) + \frac{\mu \varepsilon}{2} + \frac{1}{2}\left\| {\mathbf{Y} - \underbrace{\mathbf{W}^H \mathbf{H}\mathbf{F}(\check{\mathbf{S}}+\mathbf{E})}_{\mathbf{\tilde{\Theta}}}} \right\|_F^2+\frac{\varrho}{2}\|\mathbf{\mathcal{P}}_{\mathbf{\mathcal{T}}}(\mathbf{\Psi}_0)-\mathbf{\Psi}_0\|_F^2 + \lambda \psi_\tau(\mathbf{e}),
%\end{aligned}
\end{equation}
where $\mathbf{E}=[\mathrm{vec}^{-1}(\mathbf{e})]$. We have
\begin{equation}
\left\| {\mathbf{Y} - \mathbf{\tilde{\Theta}}} \right\|_F^2 =\mathrm{Tr}({\mathbf{Y}}^H\mathbf{Y})-\mathrm{Tr}(\mathbf{\tilde{\Theta}}^H \mathbf{Y})-\mathrm{Tr}(\mathbf{Y}^H\mathbf{\tilde{\Theta}})+ \mathrm{Tr}(\mathbf{\tilde{\Theta}}^H\mathbf{\tilde{\Theta}}).
\end{equation}
Note that
\begin{equation*}
\begin{aligned}
&\nabla_{\mathbf{E}_{i,j}}\mathrm{Tr}({\mathbf{Y}}^H\mathbf{Y})=0,\\
&\nabla_{\mathbf{E}_{i,j}}\mathrm{Tr}(\mathbf{Y}^H\mathbf{\tilde{\Theta}})
\xlongequal{(45)~in~[35]}
\left(\nabla_{\mathbf{E}_{j,i}^*}\mathrm{Tr}(\mathbf{Y}^H\mathbf{\tilde{\Theta}})\right)^H=
\left(\nabla_{\mathbf{E}_{j,i}^*}\mathrm{Tr}(\mathbf{Y}^H\mathbf{W}^H \mathbf{H}\mathbf{F}\mathbf{E})\right)^H\\
&~~~~~~~~~~~~~~~~~~\xlongequal{(233),(240),(241)~in~[35]}{[2\mathbf{Y}^H\mathbf{W}^H \mathbf{H}\mathbf{F}]}_{i,j}^H={[2\mathbf{F}^H\mathbf{H}^H \mathbf{W}\mathbf{Y}]}_{i,j},\\
&\nabla_{\mathbf{E}_{i,j}}\mathrm{Tr}(\mathbf{\tilde{\Theta}}^H\mathbf{Y})=
\nabla_{\mathbf{E}_{i,j}}\mathrm{Tr}(\mathbf{E}^H\mathbf{F}^H\mathbf{H}^H\mathbf{W}
\mathbf{Y})\xlongequal{\mathrm{Tr}(\mathbf{AB})=\mathrm{Tr}(\mathbf{BA})}\nabla_{\mathbf{E}_{i,j}}\mathrm{Tr}(\mathbf{F}^H\mathbf{H}^H\mathbf{W}
\mathbf{Y}\mathbf{E}^H)\\
&~~~~~~~~~~~~~~~~~~\xlongequal{(233),(240),(241)~in~[35]}{[2\mathbf{F}^H\mathbf{H}^H \mathbf{W}\mathbf{Y}]}_{i,j},\\
&\nabla_{\mathbf{E}_{i,j}}\mathrm{Tr}(\mathbf{\tilde{\Theta}}^H\mathbf{\tilde{\Theta}})= \nabla_{\mathbf{E}_{i,j}}\mathrm{Tr}(\mathbf{E}^H\mathbf{J}\mathbf{E})+ \nabla_{\mathbf{E}_{i,j}}\mathrm{Tr}(\mathbf{E}^H\mathbf{J}\mathbf{\check{S}})+
\nabla_{\mathbf{E}_{i,j}}\mathrm{Tr}(\mathbf{\check{S}}^H\mathbf{J}\mathbf{E})\\
&~~~~~~~~~~~~~~~~~~\xlongequal{\mathrm{Tr}(\mathbf{AB})=\mathrm{Tr}(\mathbf{BA})}
\nabla_{\mathbf{E}_{i,j}}\mathrm{Tr}(\mathbf{J}\mathbf{E}\mathbf{E}^H)+
\nabla_{\mathbf{E}_{i,j}}\mathrm{Tr}(\mathbf{J}\mathbf{\check{S}}\mathbf{E}^H)+
\nabla_{\mathbf{E}_{i,j}}\mathrm{Tr}(\mathbf{\check{S}}^H\mathbf{J}\mathbf{E})\\
&~~~~~~~~~~~~~~~~~~\xlongequal[chain~rule]{(233),(240),(241)~in~[35]}\underbrace{4\mathbf{J}\mathbf{E}}_{\nabla_{\mathbf{E}_{i,j}}\mathrm{Tr}(\mathbf{J}\mathbf{E}\mathbf{E}^H)}
+\underbrace{2\mathbf{J}\mathbf{\check{S}}}_{\nabla_{\mathbf{E}_{i,j}}\mathrm{Tr}(\mathbf{E}^H\mathbf{J}\mathbf{\check{S}})}
+\underbrace{2\mathbf{J}\mathbf{\check{S}}}_{\nabla_{\mathbf{E}_{i,j}}\mathrm{Tr}(\mathbf{\check{S}}^H\mathbf{J}\mathbf{E})}\\
&~~~~~~~~~~~~~~~~~~={[4\mathbf{J}(\check{\mathbf{S}}+\mathbf{E})]}_{i,j}, \mathbf{J}=\mathbf{F}^H\mathbf{H}^H \mathbf{W}\mathbf{W}^H \mathbf{H}\mathbf{F},
\end{aligned}
\end{equation*}
where $\mathbf{E}_{i,j}$ denotes the $(i,j)$-th element of $\mathbf{E}$. Hence
\begin{equation}\label{Derive}
\nabla_{\mathbf{e}}\zeta =\mathrm{vec}(2\mathbf{F}^H \mathbf{H}^H \mathbf{W}(\mathbf{W}^H \mathbf{H}\mathbf{F}(\check{\mathbf{S}}+\mathrm{vec}^{-1}(\mathbf{e})) - \mathbf{Y})) + \lambda \nabla_{\mathbf{e}}\psi,
\end{equation}
where the $m$-th element of $\nabla_{\mathbf{e}}\psi \in \mathbb{C}^{KM_s\times 1}$ is
\begin{equation}\label{E}
\nabla_{e_m}\psi = \frac{\sinh(|e_m|/\tau)e_m}{\cosh(|e_m|/\tau)|e_m|}.
\end{equation}
%and $\tilde{\mathbf{I}}$ is an $(K{M_s}^2\times K)$-dimensional matrix whose $(i,j)$-th elements, e.g., $i=\beta+\kappa_{\beta} M_s,j=1,2,...,K;\kappa_{\beta}=(\beta-1)K,(\beta-1)K+1,...,\beta K-1;\beta=1,2,...,M_s$, are ones and the other elements are zeros.

Then we derive $\nabla_{\mathbf{\Gamma}}\zeta$. Following the chain rule, we have
\begin{equation}\label{Chain}
\nabla_{\mathbf{\Gamma}}\zeta=2\left[\nabla_{\mathbf{\Psi}} \zeta|_{\mathbf{\Psi}=\mathbf{\Gamma}\mathbf{\Gamma}^H} \right]\mathbf{\Gamma},
\end{equation}
so the problem becomes calculating $\nabla_{\mathbf{\Psi}}\zeta$. We have
\begin{equation*}
\begin{aligned}
\|\mathbf{\mathcal{T}}(\mathds{G}(\mathbf{\Psi}_0))-\mathbf{\Psi}_0\|_F^2 &= \sum\limits_{i=-N_t+1}^{N_t-1}\sum\limits_{j=-N_r+1}^{N_r-1}\left(\sum\limits_{p-q=i}^{m-n=j}(d_{p,q}^{m,n})^2
+\kappa_{i,j}(\tilde d_{p,q}^{m,n})^2-2(\sum\limits_{p-q=i}^{m-n=j}d_{p,q}^{m,n})\tilde d_{p,q}^{m,n}\right)\\
&\xlongequal{\kappa_{i,j}\tilde d_{p,q}^{m,n}=\sum\limits_{p-q=i}^{m-n=j}d_{p,q}^{m,n}} \sum\limits_{i=-N_t+1}^{N_t-1}\sum\limits_{j=-N_r+1}^{N_r-1}\left(\sum\limits_{p-q=i}^{m-n=j}(d_{p,q}^{m,n})^2
-\frac{(\sum\limits_{p-q=i}^{m-n=j}d_{p,q}^{m,n})^2}{\kappa_{i,j}}\right).
\end{aligned}
\end{equation*}
Hence, we rewrite $\zeta(\mathbf{\Psi},\mathbf{e})$ as
\begin{equation}
\begin{aligned}
\zeta(\mathbf{\Psi},\mathbf{e})&=\frac{\mu}{2N_tN_r}\mathrm{Tr}(\mathbf{\Psi}_0) + \frac{\mu \varepsilon}{2} + \frac{1}{2}\left\| {{{\tilde{\mathbf{y}}}} - ((\check{\mathbf{S}}^T+[\mathrm{vec}^{-1}(\mathbf{e})]^T)\mathbf{F}^T \otimes \mathbf{W}^H){{\tilde{\mathbf{h}}}}} \right\|_2^2 + \lambda \psi_\tau(\mathbf{e})\\
& + \frac{\varrho}{2}\left[\sum\limits_{i=-N_t+1}^{N_t-1}\sum\limits_{j=-N_r+1}^{N_r-1}\left(\underbrace {\mathbf{m}_{i,j}^H \mathbf{m}_{i,j} - \frac{1}{\kappa_{i,j}}\mathbf{m}_{i,j}^H \mathbf{\bar{l}}_{\kappa_{i,j}} \mathbf{\bar{l}}_{\kappa_{i,j}}^H \mathbf{m}_{i,j}}_{\phi_{i,j}}\right)\right],
\end{aligned}
\end{equation}
where $\mathbf{\Psi}=\left[ {\begin{array}{*{20}{c}}
{\mathbf{\Psi}_0}&{\tilde{\mathbf{h}}}\\
{\tilde{\mathbf{h}}^H}&{\varepsilon}
\end{array}} \right]$, $\mathbf{m}_{i,j}=[d_{p_1,q}^{m_1,n},d_{p_2,q}^{m_1,n},...,d_{p_{N_t-|i|},q}^{m_1,n},d_{p_1,q}^{m_2,n},...,d_{p_{N_t-|i|},q}^{m_{N_r-|j|},n}]^T\in \mathbb{C}^{\kappa_{i,j}\times 1}$ with $m_k$ being the $k$-th smallest element among the set $\{m\}_{m-n=j}$ and $p_{\nu}$ being the $\nu$-th smallest element among the set $\{p\}_{p-q=i}$, and $\mathbf{\bar{l}}_n=[1,1,...,1]^T$ is an $n$-dimensional all one vector. After manipulation, $\zeta(\mathbf{\Psi},\mathbf{e})$ can be rewritten in a quadratic form:
\begin{equation}\label{Devitation}
\begin{aligned}
\zeta(\mathbf{\Psi},\mathbf{e})&=\underbrace{\frac{\mu}{2N_tN_r}\mathrm{Tr}(\mathbf{\Psi}_0) + \frac{\mu \varepsilon}{2}
-\frac{1}{2}\mathrm{Tr}({\tilde{\mathbf{h}}}{\tilde{\mathbf{y}}}^H \mathbf{\Theta})-\frac{1}{2}{\tilde{\mathbf{h}}}^H \mathbf{\Theta}^H {\tilde{\mathbf{y}}}}_{\langle \mathbf{A},\mathbf{\Psi}\rangle}
+\underbrace{\lambda\psi_\tau(\mathbf{e})+\frac{1}{2}{\tilde{\mathbf{y}}}^H {\tilde{\mathbf{y}}}}_{\tilde{\zeta}(\mathbf{e})}\\
&+\underbrace{\frac{\varrho}{2}\left[\sum\limits_{i=-N_t+1}^{N_t-1}\sum\limits_{j=-N_r+1}^{N_r-1}\phi_{i,j}\right]
+\frac{1}{4}{\tilde{\mathbf{h}}}^H \mathbf{\Theta}^H \mathbf{\Theta}{\tilde{\mathbf{h}}}+\frac{1}{4}\mathrm{Tr}(\mathbf{\Theta}^H \mathbf{\Theta}{\tilde{\mathbf{h}}}{\tilde{\mathbf{h}}}^H)}_{\langle \mathbf{\Psi},\mathcal{Q}(\mathbf{\Psi}) \rangle/2}
\end{aligned}
\end{equation}
where $\tilde{\zeta}(\mathbf{e})$ is a function that depends on $\mathbf{e}$; $\mathbf{A} \in \mathbb{C}^{(N_tN_r+1)\times(N_tN_r+1)}$ and $\mathcal{Q}(\mathbf{\Psi}) \in \mathbb{C}^{(N_tN_r+1)\times(N_tN_r+1)}$ can be respectively computed by
\begin{equation}\label{A}
\mathbf{A}~~= \frac{1}{2}\left[ \begin{array}{cc}
                                 \frac{\mu}{N_tN_r}\mathbf{I}_{N_tN_r} & -\mathbf{\Theta}^H\tilde{\mathbf{y}} \\
                                 -\tilde{\mathbf{y}}^H\mathbf{\Theta} & \mu
                               \end{array}
\right],
\end{equation}
\begin{equation}\label{Q}
\mathcal{Q}(\mathbf{\Psi}) = \left[ \begin{array}{cc}
                                 \mathbf{\Xi}(\mathbf{\Psi}_0) & \mathbf{\Theta}^H\mathbf{\Theta}\tilde{\mathbf{h}}/2 \\
                                 \tilde{\mathbf{h}}^H\mathbf{\Theta}^H\mathbf{\Theta}/2& 0
                               \end{array}
\right],
\end{equation}
%\begin{equation}
%\mathbf{Z}=({{\tilde{\mathbf{y}}}} - (\mathbf{E}^T \mathbf{F}^T \otimes \mathbf{W}^H){{\tilde{\mathbf{h}}}})(\mathbf{E}^T \mathbf{F}^T \otimes \mathbf{W}^H),
%\end{equation}
with
\begin{eqnarray}
\mathbf{\Xi}(\mathbf{\Psi}_0)&=&\varrho\sum\limits_{j=-Nr+1}^{N_r-1}\sum\limits_{i=-N_t+1}^{N_t-1}
\varphi(\mathbf{m}_{i,j}-\frac{1}{\kappa_{i,j}}\mathbf{\bar{l}}_{\kappa_{i,j}}\mathbf{\bar{l}}_{\kappa_{i,j}}^H \mathbf{m}_{i,j}),
%\nabla_{\mathbf{m}_{i,j}}\phi_{i,j}&=&2(\mathbf{m}_{i,j}-\frac{1}{\kappa_{i,j}}\mathbf{\bar{l}}_{\kappa_{i,j}}\mathbf{\bar{l}}_{\kappa_{i,j}}^H \mathbf{m}_{i,j}),
\end{eqnarray}
and $\varphi(\mathbf{m}_{i,j}-\frac{1}{\kappa_{i,j}}\mathbf{\bar{l}}_{\kappa_{i,j}}\mathbf{\bar{l}}_{\kappa_{i,j}}^H \mathbf{m}_{i,j})$
outputs an $N_tN_r\times N_tN_r$ matrix, which can be divided into $N_r \times N_r$ blocks
$\mathbf{\hat D}_{m,n}$ with the $(p,q)$-th element of $\mathbf{\hat D}_{m,n}$ being $\hat{d}_{p,q}^{m,n}$,
$m,n=1,2,...,N_r$, $p,q=1,2,...,N_t$, and
$[\hat d_{p_1,q}^{m_1,n},\hat d_{p_2,q}^{m_1,n},...,\hat d_{p_{N_t-|i|},q}^{m_1,n},\hat d_{p_1,q}^{m_2,n},...,\hat d_{p_{N_t-|i|},q}^{m_{N_r-|j|},n}]^T = \mathbf{m}_{i,j}-\frac{1}{\kappa_{i,j}}\mathbf{\bar{l}}_{\kappa_{i,j}}\mathbf{\bar{l}}_{\kappa_{i,j}}^H \mathbf{m}_{i,j}$,
and the rest of the elements are zeros.
%whose element $\hat{d}_{p,q}^{m,n}, p-q=i, m-n=j,$ is the $((\jmath(m)-1)(N_r-|j|)+\wp(p))$-th entry of the input vector $\nabla_{\mathbf{m}_{i,j}}\phi_{i,j}$, and the rest of the elements are zeros. $\jmath(m) \in O(\{m\}_{m-n=j})$ is the index of input $m$, where $O(\{m\}_{m-n=j})=\{\jmath(m)\}_{m-n=j}$ index the input set $\{m\}_{m-n=j}$ according to the order of $m$ from the smallest to the largest and starts numbering from one. Similarly to $\jmath(m)$, $\wp(p)$ is the index of input $p$.

From (\ref{Devitation}),
\begin{align}
&\nabla_{\mathbf{\Psi}_{r,z}}(\langle\mathbf{\Psi},\mathcal{Q}(\mathbf{\Psi})\rangle/2)\\
&=\left\{\begin{gathered}
\nabla_{d^{m,n}_{p,q}}\frac{\varrho}{2}\phi_{i,j}\xlongequal[chain~rule]{(233),(236),(237)~in~[35]} 2\varrho(d^{m,n}_{p,q}-\frac{1}{\kappa_{i,j}}\mathbf{\bar{l}}_{\kappa_{i,j}}^H \mathbf{m}_{i,j}), r,z=1,...,N_tN_r, d^{m,n}_{p,q} \in \mathbf{m}_{i,j}, \hfill \\
\nabla_{{\tilde{\mathbf{h}}}_r}\frac{1}{4}\left({\tilde{\mathbf{h}}}^H \mathbf{\Theta}^H \mathbf{\Theta}{\tilde{\mathbf{h}}}+\mathrm{Tr}(\mathbf{\Theta}^H \mathbf{\Theta}{\tilde{\mathbf{h}}}{\tilde{\mathbf{h}}}^H)\right) ={[\mathbf{\Theta}^H\mathbf{\Theta}\tilde{\mathbf{h}}]}_r,~~~~~~~r=1,...,N_tN_r, z=N_tN_r+1, \hfill \\
\nabla_{{{\tilde{\mathbf{h}}}^H}_z}\frac{1}{4}\left({\tilde{\mathbf{h}}}^H \mathbf{\Theta}^H \mathbf{\Theta}{\tilde{\mathbf{h}}}+\mathrm{Tr}(\mathbf{\Theta}^H \mathbf{\Theta}{\tilde{\mathbf{h}}}{\tilde{\mathbf{h}}}^H)\right) ={[\tilde{\mathbf{h}}^H\mathbf{\Theta}^H\mathbf{\Theta}]}_z,~~~~~z=1,...,N_tN_r,r=N_tN_r+1, \hfill \\
\nabla_{\varepsilon}\left(\langle \mathbf{\Psi},\mathcal{Q}(\mathbf{\Psi}) \rangle/2\right)=0,~~~~~~~~~~~~~~~~~~~~~~~~~~~~~~~~~~~~~~~~r=z=N_tN_r+1, \hfill
\end{gathered} \right\},
\end{align}
where $\mathbf{\Psi}_{r,z}$ denotes the $(r,z)$-th element of $\mathbf{\Psi}$.

Hence
\begin{equation}\label{Deri}
\nabla_{\mathbf{\Psi}}\zeta = 2\mathcal{Q}(\mathbf{\Psi})+2\mathbf{A}.
%\nabla_{\mathbf{\Psi}}\zeta \xlongequal{(6)~in~[35]} \mathcal{Q}(\mathbf{\Psi})+\mathbf{A}.
\end{equation}
Plugging (\ref{Deri}) into (\ref{Chain}), $\nabla_{\mathbf{\Gamma}}\zeta$ can be obtained.

%\section*{Acknowledgment}

\ifCLASSOPTIONcaptionsoff
  \newpage
\fi

\end{document}